\documentclass[review, brazil, english]{elsarticle}
\usepackage{subcaption}
\biboptions{numbers,sort&compress,semicolon}

\makeatletter
\def\ps@pprintTitle{%
  \let\@oddhead\@empty
  \let\@evenhead\@empty
  \let\@oddfoot\@empty
  \let\@evenfoot\@oddfoot
}
\makeatother

\UseRawInputEncoding
\usepackage[titletoc]{appendix}
\usepackage{supertabular}
\usepackage{lineno,hyperref} %
\usepackage{siunitx}
\usepackage{soul}
\usepackage{listings}
\usepackage{setspace,lipsum}
\usepackage{comment}
\usepackage{multirow}
\usepackage{xcolor}
\usepackage{lscape}
\usepackage{bm}

\usepackage[T1]{fontenc}
\hypersetup{pdfauthor=author}
\usepackage{tikz}
\usetikzlibrary{shapes.multipart}
\usepackage{xcolor}
\usepackage{framed}
\usepackage{pdflscape}
\usepackage{appendix}
\usepackage{booktabs}
\usepackage{tcolorbox}
\usepackage{todonotes}

\colorlet{punct}{red!60!black}
\definecolor{background}{HTML}{EEEEEE}
\definecolor{delim}{RGB}{20,105,176}
\colorlet{numb}{magenta!60!black}

\newcommand{\Reu}{\mathrm{Re}}

\newcommand{\vm}[1]{\bm{#1}}

\newcommand{\dd}[2]{\frac{\partial #1}{\partial #2}}

\newcommand{\pp}[1]{\left(#1\right)}

\modulolinenumbers[5]

\usepackage[skip=0pt, margin = 0pt, font=normal]{subcaption}
\usepackage{bm,amsmath,amssymb}
\usepackage{tikz}
    \usetikzlibrary{arrows, positioning, shapes}
\journal{Chemical Engineering Research and Design}

\lstdefinelanguage{json}{
	basicstyle=\small \ttfamily,
	numberstyle=\scriptsize,
	stepnumber=1,
	numbersep=8pt,
	showstringspaces=false,
	breaklines=true,
	frame=lines,
	backgroundcolor=\color{background},
	literate=
	*{0}{{{\color{numb}0}}}{1}
	{1}{{{\color{numb}1}}}{1}
	{2}{{{\color{numb}2}}}{1}
	{3}{{{\color{numb}3}}}{1}
	{4}{{{\color{numb}4}}}{1}
	{5}{{{\color{numb}5}}}{1}
	{6}{{{\color{numb}6}}}{1}
	{7}{{{\color{numb}7}}}{1}
	{8}{{{\color{numb}8}}}{1}
	{9}{{{\color{numb}9}}}{1}
	{:}{{{\color{punct}{:}}}}{1}
	{,}{{{\color{punct}{,}}}}{1}
	{\{}{{{\color{delim}{\{}}}}{1}
	{\}}{{{\color{delim}{\}}}}}{1}
	{[}{{{\color{delim}{[}}}}{1}
	{]}{{{\color{delim}{]}}}}{1},
}

\begin{document}

\begin{frontmatter}

\title{Surrogate Model for Heat Transfer Prediction in Impinging Jet Arrays using Dynamic Inlet/Outlet and Flow Rate Control}

%% or include affiliations in footnotes:
\author[chaos,poly]{Mikael Vaillant}
\author[chaos,poly]{Victor Oliveira Ferreira}
\author[chaos,poly]{Wiebke Mainville}
\author[cnrc]{Jean-Michel Lamarre}
\author[cnrc]{Vincent Raymond}
\author[poly]{Moncef Chioua}
\author[chaos,poly]{Bruno Blais\corref{mycorrespondingauthor}}
\cortext[mycorrespondingauthor]{Corresponding author.}
\ead{bruno.blais@polymtl.ca}

\address[chaos]{Chemical Engineering High-Performance Analysis Optimization and Simulation (CHAOS) Laboratory, Department of Chemical Engineering, Polytechnique Montr\'eal, 2500 chemin de Polytechnique, QC, Canada, H3T 1J4}

\address[cnrc]{National Research Council Canada, Boucherville, QC, Canada, J4B 6Y4 }

\address[poly]{Department of Chemical Engineering, Polytechnique Montr\'eal, PO Box 6079, Stn Centre-Ville, Montr\'eal, QC, Canada, H3C 3A7.}

\begin{abstract}
%*****************************
% Abstract
%*****************************

% This study presents a surrogate model designed to predict the Nusselt number distribution in an enclosed impinging jet arrays, where each jet function independently and where jets can be transformed from inlets to outlets, leading to a vast number of possible flow arrangements. While computational fluid dynamics (CFD) simulations can model heat transfer with high fidelity, their cost prohibits real-time application such as model-based temperature control. To address this, we generate a CNN-based surrogate model that can predict the Nusselt distribution in real time. We train it with data from implicit large eddy computational fluid dynamics simulations (Re < 2,000). We train two distinct models, one for a five by one array of jets (83 simulations) and one for a three by three array of jets (100 simulations). We introduce a method to extrapolate predictions to higher Reynolds numbers (Re < 10,000) using a correlation-based scaling. The surrogate models achieve high accuracy, with a normalized mean average error below 2\% on validation data for the five by one surrogate model and 0.6\% for the three by three surrogate model, and their predictions are validated experimentally. This work provides a foundation for model-based control strategies in advanced thermal management applications.

This study presents a surrogate model designed to predict the Nusselt number distribution in an enclosed impinging jet arrays, where each jet function independently and where jets can be transformed from inlets to outlets. While computational fluid dynamics (CFD) simulations can model heat transfer with high fidelity, their cost prohibits real-time application such as model-based temperature control. To address this, we generate a CNN-based surrogate model that predicts the Nusselt distribution in real time. We train it with data from implicit large-eddy computational fluid dynamics simulations (Re < 2,000). We train two distinct models, one for a five by one array of jets (83 simulations) and one for a three by three array of jets (100 simulations). We introduce a method to extrapolate predictions to higher Reynolds numbers (Re < 10,000) using a correlation-based scaling. The surrogate models achieve high accuracy, with a normalized mean average error below 2\% on validation data for the five by one surrogate model and 0.6\% for the three by three surrogate model, and their predictions are validated experimentally. This work provides a foundation for model-based control strategies in advanced thermal management applications.
\end{abstract}


\end{frontmatter}

% \linenumbers

%**********************************************************
% Introduction
%**********************************************************

\section{Introduction}
%*****************************
% Introduction
%*****************************

Impinging jets are key in thermal management applications such as part manufacturing, gas turbine cooling, electric vehicle battery cooling \cite{20244217219667}, and electronic cooling systems \cite{chiriac2002numerical}. Their prevalence in such technologies results from their capacity to generate localized high Nusselt numbers that can be three times greater than what is obtained for parallel flow for the same flow rate \cite{zuckerman2006jet}. While single impinging jets can provide much higher Nusselt number at the center of the jet, impinging jet arrays increase the surface coverage on the impinged surface and produce a more homogeneous Nusselt number profile \cite{Zhao23022005}.

Aiming to take advantage of the jet array's coverage as well as the impinging jets' localized high Nusselt number, \citet{Lamarre_2023} developed a thermal management device to control the temperature of a whole surface as a function of time and space. The thermal management device is designed to suit any technology that could require spatial and temporal temperature control. A potential application is injection molds cooling systems, where different technologies such as the one proposed by \citet{hopmann2020development} have been proposed for local temperature control. Other promising candidates include thermal photovoltaic systems \cite{20243817071309} and automotive batteries \cite{Fu_2022}, where temperature control could improve performance and efficiency. Examples of such a device are shown in Figures \ref{fig:5x1_geometry} and \ref{fig:3x3_geometry}. The device comprises a cavity with jet nozzles pointing to a temperature-controlled surface. In this thermal management device, each jet has its own independent flow rate and can act as either an inlet, an outlet, or be shut, dynamically switching between states. In this paper, we refer to these thermal management devices as active cooling systems. We also define the inlet/outlet/shut status of a jet as a \textit{state} and the configuration of multiple jets (meaning the state of every jet within an array of jets) as an \textit{arrangement}.

Other technologies using impinging jets have recently shown potential for surface temperature control. For instance, a technology developed by \citet{fujimori2024dns} uses groups of pulsating jets to adjust the heat load profile in space of an array of jets. However, this technology, just as the one developed by Lamarre and Raymond, requires a closed-loop surface temperature control strategy to address time-varying heat loads. This type of closed-loop surface temperature control has already been successfully implemented in other works, such as the work of \citet{salavatidezfouli2023deep, salavatidezfouli2024predictive}, which uses deep reinforcement techniques to control the surface temperature of a single pulsating jet. However, such closed loop temperature control strategies have yet to be applied to an enclosed array of impinging jets.

In the technology developed by Lamarre and Raymond, a closed loop temperature control strategy modulates the flow rates of the jets, but also dynamically changes the arrangement (i.e, change which jet acts as an inlet or outlet). Developing a temperature control strategy rapidly increases in complexity with the number of jets since increasing the number of jets leads to a greater number of possible arrangements. While this allows for finer spatial and temporal control of the Nusselt number distribution, it also increases the complexity of jet-jet interactions \cite{weigand2009multiple} and makes the evaluation of the performance of control strategies under varying thermal loads more difficult. For illustration, a five by five array of jets containing only the inlet/outlet/shut states could lead to $\approx3^{25}$ arrangements ($\approx$ 1 trillion). Even if the arrangement symmetries and no-flow conditions are removed, evaluating the performance of the control strategies for the remaining cases is, at the very least, challenging.

% Start of the litterature review
Therefore, a model capable of evaluating efficiently the Nusselt number distribution within the active cooling system facilitates the development and the evaluation of control strategies. Additionally, this model enables model-based control which can, in the end, provide dynamic flow rate and arrangement modulation. As highlighted in several reviews \cite{weigand2009multiple, barbosa2023convection, uddin2024heat}, the evaluation of the Nusselt number distribution produced by impinging jets has traditionally been based on empirical correlations and computational fluid dynamics (CFD) simulations. 

For instance, correlations proposed by authors such as \citet{martin1977heat}, \citet{jambunathan1992review} or \citet{katti2008experimental} relate the Nusselt number to key dimensionless parameters: the Reynolds number, Prandtl number, jet-to-surface spacing ratio (H/D), and the radial distance from the center of the jet. In single-jet configurations, incorporating the radial distance from the center of the jet provides valuable spatial resolution of the heat transfer profile. However, in the case of jet array correlations, such as the one proposed by \citet{jet_array_correlation}, only the average Nusselt number above each jet is typically considered. More importantly, such correlations are generally developed for a fixed arrangement where all jets are active, limiting their applicability to different configurations such as the ones found in the active cooling system described by \citet{Lamarre_2023}. This is where CFD becomes particularly valuable, as it enables the precise simulation and analysis of cases using different jet arrangements and jet flow rates.

Numerous studies have employed CFD to predict heat transfer in impinging jet arrays, as discussed in the reviews by \citet{weigand2009multiple} and \citet{barbosa2023convection}. While the focus has been placed on single jet impingement \cite{plant2023review, uddin2024heat}, jet array simulation has also received some attention, namely to address the jet-jet interactions and the effect of key geometrical (e.g., jet-to-jet distance, jet-to-plate distance, or jet pattern) and flow parameters (e.g., Reynolds number, Prandtl number) on these interactions \cite{weigand2009multiple, barbosa2023convection}. A parametric study on jet-to-jet distance and cross-flow by \citet{otero2021high} showed that jet-jet interactions can reduce the Nusselt number. This jet-jet interaction must then be considered when developing a predictive heat transfer model because a change in the flow rate of a jet or the addition of an outlet can affect the Nusselt number profile over the surface that is being cooled (or heated). CFD simulations can capture the effect of jet-jet interactions on the Nusselt number distribution. However, their computational cost remains prohibitive for real-time applications such as temperature control. This limitation is further amplified by the high Reynolds numbers typically observed in such systems, which significantly increase the computational cost. 

To leverage the accuracy of CFD simulations while reducing computational cost, several studies have recently shifted towards the use of surrogate models. These models are typically artificial neural networks (ANNs) trained on CFD-generated datasets to predict the Nusselt number. For instance, \citet{singh2021numerical} used simulation data to train an ANN aimed at identifying optimal geometrical parameters to enhance heat transfer. Similarly, \citet{fawaz2024artificial} developed an ANN to predict the Nusselt number in impinging swirling flows. More recent studies such as the one by \citet{20252218537271} have used machine learning to predict the Nusselt number in impinging jet arrays. The use of surrogate models is not limited to impinging jets, other fluid-related problems such as mixing have also employed this strategy extensively \cite{bibeau2023artificial}.

Existing studies have focused primarily on fixed configurations. To the best of our knowledge, no surrogate model has been developed for impinging jet arrays with dynamic inlet and outlet configurations. Unlike direct CFD simulations, surrogate models enable rapid evaluation of the Nusselt number for a given configuration and could, in future work, be used within model-based closed-loop temperature control strategies for jet arrays.

% These works highlight the potential of surrogate models trained on CFD simulations to predict the Nusselt number on the impinged surface of the active cooling system. Unlike direct CFD simulations, which become impractical for real-time prediction across numerous arrangements, this approach allows efficient estimation for any jet configuration. To the best of our knowledge, no existing work has developed surrogate models specifically for impinging jet arrays with dynamic inlet/outlet configurations. This surrogate model will also enable the development of model-based closed-loop temperature control strategies for jet arrays, which represents a significant contribution given that previous approaches have been limited to single jets or systems that do not involve impinging jets.

In this work, we propose a surrogate model to approximate the Nusselt number on an impinged wall for any arrangement to allow extensive control strategy testing while also providing a tool for future model-based control. We focus on two distinct active cooling systems shown in Figures \ref{fig:5x1_geometry} and \ref{fig:3x3_geometry}: a five by one and a three by three jet arrangement, for which separate surrogate models are developed. These surrogate models use a convolutional neural network (CNN) and are trained using time-averaged CFD simulation results obtained using implicit large-eddy simulations. While the neural network is trained to predict Nusselt number profiles of various arrangements at inlet Reynolds numbers under 2,000, we also provide a method to extrapolate the predictions to equivalent arrangements at higher Reynolds number (Re $\leq$ 10,000) by adapting a correlation proposed by \citet{martin1977heat}. 

To present the development of the surrogate model, we begin by detailing the geometry of the active cooling system. Next, we introduce key dimensionless parameters such as the jet-jet distance, cavity to jet ratio and number of jets in the system. We then establish the framework we use to generate the CFD simulation dataset, along with the CFD software used. The equations, the simulations parameters (e.g., total simulation time, total averaging time for the Nusselt number), the boundary conditions and other important dimensionless parameters such as the Nusselt and Prandtl are presented. We subsequently perform a mesh sensitivity analysis to assess the mesh independence of the simulation results. This is followed by an explanation of the sampling strategy for the simulated jet arrangements and how we divided our training into different subsets for testing and validation of our surrogate model. Following this, we describe the simulation post-processing steps to train the convolutional neural network. We then present the architecture of the convolutional neural network before finally introducing the methodology for prediction extrapolation to higher Reynolds numbers.

We evaluate the performance of the active cooling system using the root mean squared error (RMSE), the normalized mean average error (NMAE), the average maximum error and the average top 10\% error. We also compare the average Nusselt number of the simulated and predicted Nusselt number of the validation dataset. We subsequently present the error between the scaled predictions at a Reynolds number of 10,000 and simulated results of the corresponding configuration. Finally, we compare simulation results using the surrogate models with experimental results and demonstrate the good agreement, therefore validating the approach.

\begin{tcolorbox}[
  colback=white,
  colframe=black,
  boxrule=0.8pt,
  left=6pt,
  right=6pt,
  top=6pt,
  bottom=6pt
]
\textbf{Nomenclature}

\vspace{0.5em}

\begin{tabular}{@{}ll@{\hspace{0.5cm}}ll@{}}
$C_p$ & Specific heat capacity, J$\cdot$kg$^{-1}$$\cdot$K$^{-1}$ & $H_p$  & Plate thickness, m  \\
$D$  & Inlet diameter, m & $t^*$  & Dimensionless time \\
$H$  & Cavity height, m & $T$ & Temperature, K  \\
$k$  & Thermal conductivity, W$\cdot$m$^{-1}\cdot$K$^{-1}$ & $\textbf{u}$ & fluid velocity m$\cdot$s$^{-1}$ \\
$l$  & Cavity length, m & $U$ & Inlet velocity, m$\cdot$s$^{-1}$ \\
$L$  & Cavity depth, m & $S$  & Jet-jet distance, m  \\
Nu  & Nusselt number & \multicolumn{2}{l}{\textit{Greek letters}} \\
$p$ & Pressure, Pa & $\mu$  & Dynamic viscosity Pa$\cdot$s \\
$\textrm{Pr}$  & Prandtl number & $\nu$  & Kinematic viscosity m$^2$$\cdot$s \\
$\textrm{Re}$   & Reynolds number & $\rho$  & fluid density kg$\cdot$m$^{-3}$ \\
\end{tabular}
\end{tcolorbox}

%***********************************************
% Geometry
%***********************************************

\section{Geometry of the active cooling device}
\begin{figure}[ht!]
    \centering
    \begin{subfigure}[b]{0.95\linewidth}
        \centering
        \includegraphics[width=\linewidth]{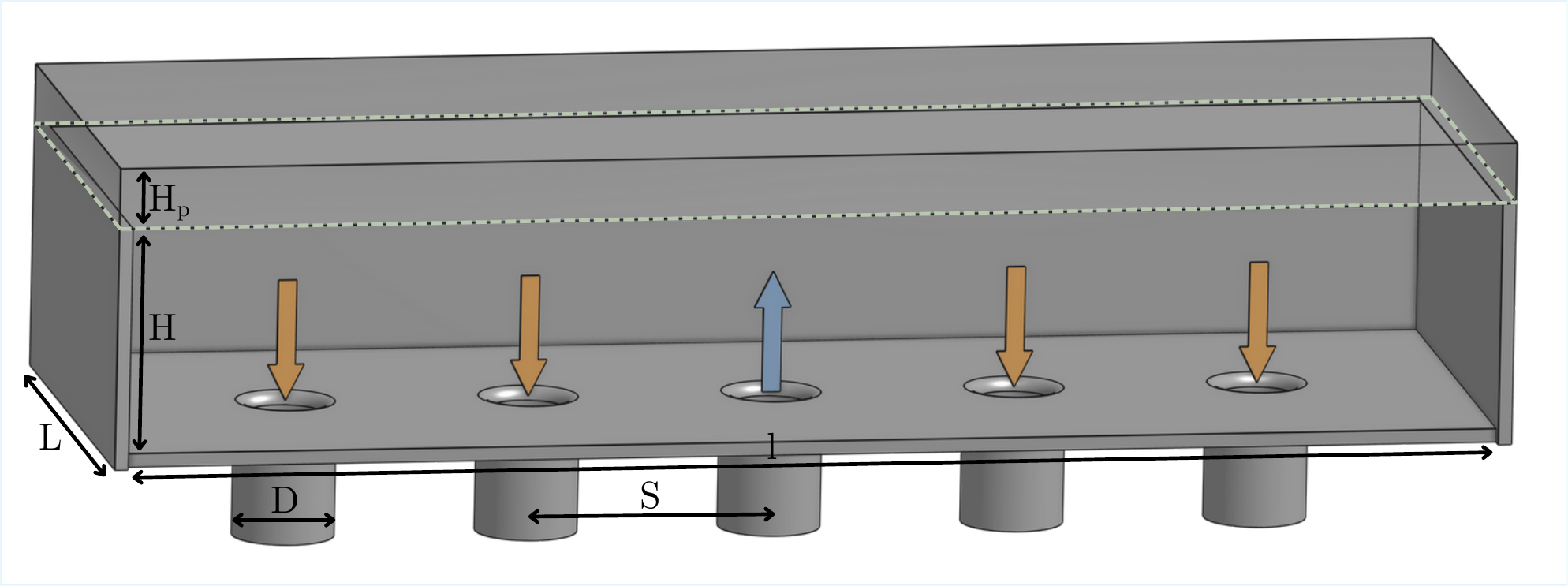}
        \caption{Five by one jet active cooling system with one inlet at position 2 and four outlets at positions 0, 1, 3, and 4.}
        \label{fig:5x1_geometry}
    \end{subfigure}
    
    \vspace{0.5cm} % Optional spacing between subfigures

    \begin{subfigure}[b]{0.9\linewidth}
        \centering
        \includegraphics[width=\linewidth]{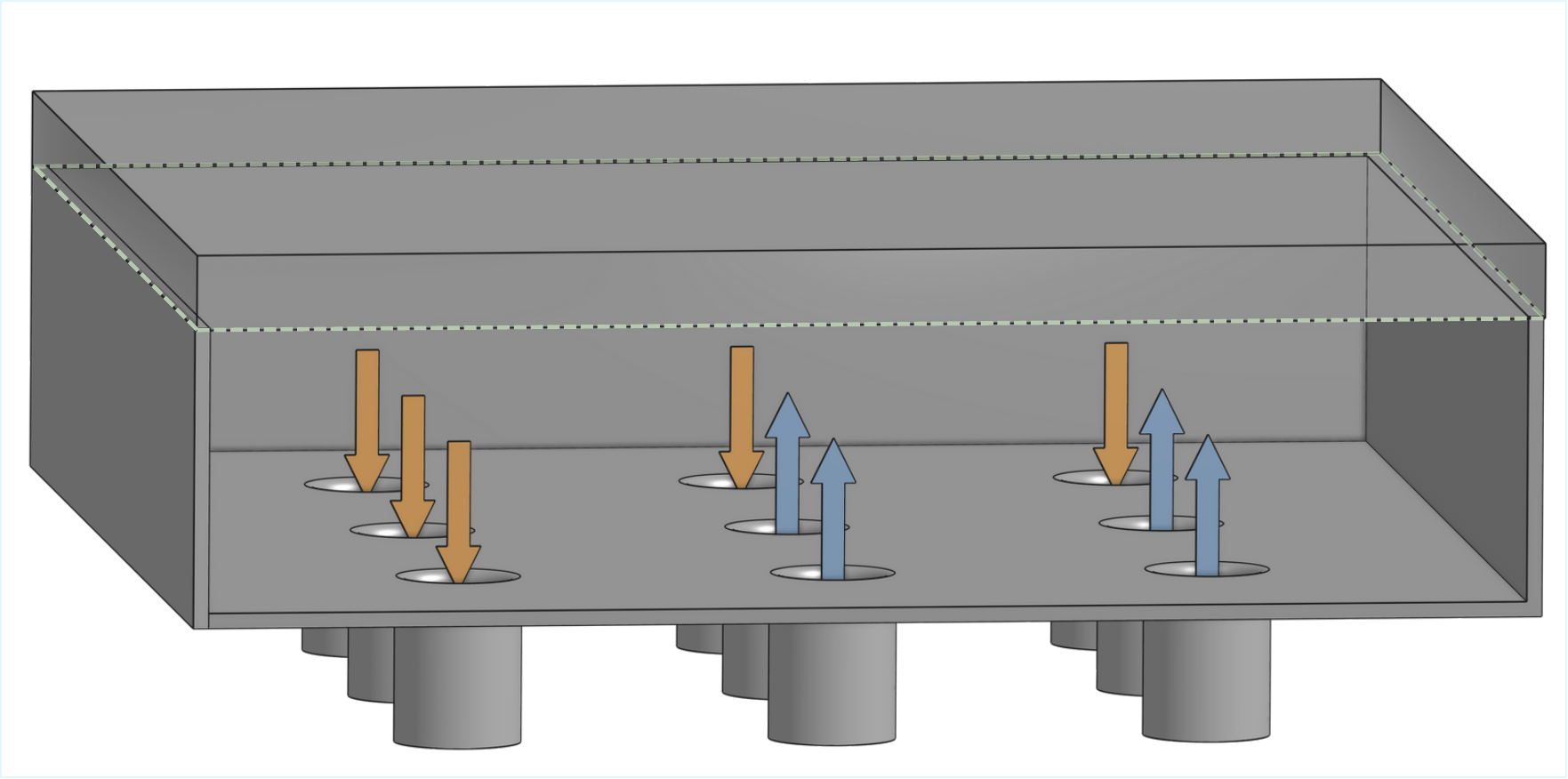}
        \caption{Three by three jet active cooling system with inlets at positions 4, 5, 7, and 8, and outlets at positions 0, 1, 2, 3, and 6.}
        \label{fig:3x3_geometry}
    \end{subfigure}
    
    \caption{Geometries of the two jet active cooling systems: (a) Five by one configuration and (b) Three by three configuration. In these systems, every inlet can be dynamically changed into an outlet and every outlet can be changed into an inlet. Any nozzle can also be shut. The surrogate models estimate the Nusselt distribution on the surface outlined by the green dashed line.}
    \label{fig:cooling_geometries}
\end{figure}

Two active cooling systems are illustrated in Figures \ref{fig:5x1_geometry} and \ref{fig:3x3_geometry}. They consist of an enclosed cavity in which the fluid can either enter or exit through jet ports. The first system contains five jets (arranged in a five by one configuration) while the second contains nine jets (arranged in a three by three configuration).

\begin{table}[ht!]
\caption{Geometric parameters of the five by one and three by three active-cooling systems. All dimensions are normalized by the inlet diameter $D$, where $D$ equals 6.35 mm in the experimental setup.}
\centering
\begin{tabular}{@{}lcc@{}}
\toprule
\textbf{Geometric parameters}    & \textbf{5 by 1} & \textbf{3 by 3} \\ \midrule
Length/Diameter ($l/D$)  [-] & 24 & 18  \\
Height/Diameter ($H/D$) [-] & 6  & 6 \\ 
Depth/Diameter ($L/D$) [-] & 6 & 18 \\
Jet-jet distance/Diameter ($S/D$)  [-] & 4 & 5 \\
Plate thickness/Diameter ($H_p/D$) [-] & 0.5 & 0.5 \\ \bottomrule 
\end{tabular}
\end{table}

We train a separate surrogate model for each of the two active cooling systems. Both surrogate models are designed to predict the Nusselt number at the bottom surface of the impingement surface, defined as the region bounded by the dashed line in Figures \ref{fig:5x1_geometry} and \ref{fig:3x3_geometry}. The first system, based on a five by one jet array, is used for mesh sensitivity analysis, experimental validation and model validation, as it corresponds to the experimental setup in our laboratory. Furthermore, the smaller physical size of the system allows for the execution of simulations with higher mesh density and greater Reynolds numbers without excessive computational cost. In contrast, the three by three configuration illustrates the scalability of the proposed methodology to systems with a larger number of jets. 

%**********************************************************
% FEM
%**********************************************************
\section{Methodology for the CFD simulations}

\subsection{Mass, momentum and heat transfer equations}

We perform all the simulations using the open-source CFD software Lethe \cite{lethe1.0} using version 1.0. This finite element software solves the incompressible Navier-Stokes equations:  

\begin{subequations}
\begin{align}
    \nabla \cdot \vm{u} &= 0    \label{eq::single_phase_ns_continuity}\\
    \dd{\vm{u}}{t} + \pp{\vm{u} \cdot \nabla} \vm{u} &= -\nabla p^*  + \nu \nabla^2  \vm{u}
    \label{eq::single_phase_ns_momentum}
\end{align}
\end{subequations}
where $\vm{u}$ is the velocity, $p^*$ the pressure $p$ divided by the density $\rho$ ($p^*=\frac{p}{\rho}$) and $\nu$ is the kinematic viscosity. Equations \ref{eq::single_phase_ns_continuity} and \ref{eq::single_phase_ns_momentum} are solved with the Finite Element Method (FEM) using a PSPG/SUPG stabilized formulation to obtain the velocity profile $\vm{u}$ and the pressure field $p^*$ \cite{tezduyar1992}. At higher Reynolds numbers, an implicit large-eddy simulation (ILES) strategy is used where the stabilization acts as a subgrid-scale model\cite{grinstein2007implicit}. This strategy has been extensively validated on complex turbulent flow benchmarks such as the flow over periodic hills, the turbulent Taylor-Couette flow, and the turbulent Taylor-Green vortex. \cite{saavedra2024implicit, PrietoSaavedra2024}

We obtain the fluid temperature by solving the enthalpy conservation for an incompressible flow: 

\begin{align} \label{eq::enthalpy_conservation}
    \rho c_\mathrm{p}\left[\dd{T}{t} + (\vm{u} \cdot \nabla)T \right]- \nabla \cdot (k\nabla T) = 0
\end{align}
where $\rho$ is the density, $c_p$ is the specific heat capacity, $k$ is the thermal conductivity and $T$ is the temperature. In this work, we assume that the density, specific heat capacity and thermal conductivity are constant.

In all simulations, the velocity and pressure fields are approximated using linear hexahedral elements with second-order accuracy (Q1Q1), while the temperature field is approximated using quadratic hexahedral elements with third-order accuracy (Q2).

\subsection{Dimensionless parameters}

We introduce four key dimensionless parameters. First, for the jets, we define the Reynolds number as:
\begin{equation}\label{eq:reynolds}
    \Reu = \frac{UD}{\nu}
\end{equation}

\noindent where $U$ is the average velocity at the inlet, $D$ is the inlet diameter and $\nu$ is the kinematic viscosity of the fluid.

We then define the Nusselt number as:
\begin{equation} \label{eq:nusselt}
    \textrm{Nu} = \frac{hD}{k}
\end{equation}

\noindent where $h$ is the heat transfer coefficient at the impinged wall, $D$ is the inlet diameter and k is the thermal conductivity. 

Next, we define the Prandtl number as:

\begin{equation}
    \textrm{Pr} = \frac{c_p\mu}{k}
\end{equation}

\noindent where $c_p$ is the heat capacity of the fluid and $\mu$ is the dynamic viscosity of the fluid. All the results presented in this article use the Prandtl number of air which is approximately 0.72.

Finally, we use the dimensionless time $t^*$ defined as:

\begin{equation}\label{eq:dimensionless_time}
    t^* = \frac{Ut}{D}
\end{equation}

Where $t$ is the simulation time. 

\subsection{Simulation setup}

We perform the low Reynolds number simulations using the parameters values provided in Table~\ref{tab:simulation_parameters}. For each simulation, we use the definition of the dimensionless time in Equation \ref{eq:dimensionless_time}. Each simulation models an interval of $t^*=60$. $t^* < 40$ allows the flow to reach a pseudo-steady state. We then start time-averaging the Nusselt number (Eq.~\ref{eq:nusselt}) at the boundary for $t^* \geq 40$. We determine the start time and the total averaging time through a time-independence study. We conduct this study using the five by one active cooling system, which consists of a single outlet located at position 0 and two inlets located at positions 3 and 4, each operating at a Reynolds number of 2,000.

\begin{table}[ht!]
\caption{Low-Reynolds simulation parameters.}
\centering
\begin{tabular}{@{}lc@{}}
\toprule
\textbf{Simulation Parameter}    & \textbf{Value (training/validation)} \\ \midrule
% Inlet simulation diameter (m)\commentText{to be revised}& 1 \\
Total time $t^*$ [-] & 60 \\
Maximum inlet Reynolds number [-]  & 2,000 / 10,000  \\
Prandtl number [-]       & 0.72  \\
Impinged wall temperature [$^{\circ}$C] & 50 \\
Inlet temperature [$^{\circ}$C] & 20 \\
Initial temperature [$^{\circ}$C] & 20 \\ \bottomrule 
\end{tabular}
\label{tab:simulation_parameters}
\end{table}

For the Navier-Stokes equations, we use a Dirichlet boundary conditions for the inlets and the walls:

\begin{equation} \label{eq:dirichlet}
    \boldsymbol{u} = f(\boldsymbol{x}) \quad \text{for } \boldsymbol{x} \in \partial \Omega_D
\end{equation}

\noindent where $\boldsymbol{u}$ is the velocity at the inlet boundary $ \partial \Omega_D$ and $f(\boldsymbol{x})$ is a function describing the velocity at this boundary. For jet inlets, we define $f(\boldsymbol{x})$ with a parabolic velocity profile whose average matches the desired velocity. The choice of this desired velocity is adjusted based on the inlet Reynolds number. For walls, we use a no-slip boundary conditions, where $f(\boldsymbol{x})$ is set to zero. The treatment of outlet boundaries follows the approach proposed by \citet{Arndt_outflow_bc} to prevent flow re-entry into the domain. 

For the enthalpy equation, we use a Dirichlet boundary conditions for both the inlets and the impingement wall. In the case of the enthalpy equation, the Dirichlet boundary condition is defined as:

\begin{equation}
    T = g(\boldsymbol{x}) \quad \text{for } \boldsymbol{x} \in \partial \Omega_D
\end{equation}

where $g(\boldsymbol{x})$ is the function describing the temperature on the boundary $\partial \Omega_D$. The temperature $T$ is set to 50$^{\circ}$C on the impinged surface and to 20$^{\circ}$C at the inlets. 

No-flux Neumann boundary conditions (insulated) are applied to both the walls and the outlets, and are defined as:

% \begin{equation}
%     \frac{\partial T}{\partial \boldsymbol{n}} = 0
% \end{equation}

\begin{equation}
    \nabla T \cdot\boldsymbol{n}= 0
\end{equation}

\subsection{Mesh convergence analysis}

\begin{figure}[ht]
  \centering
  \includegraphics[angle=270, scale=1, trim=0cm 1cm 13cm 1cm, clip, width=1\textwidth]{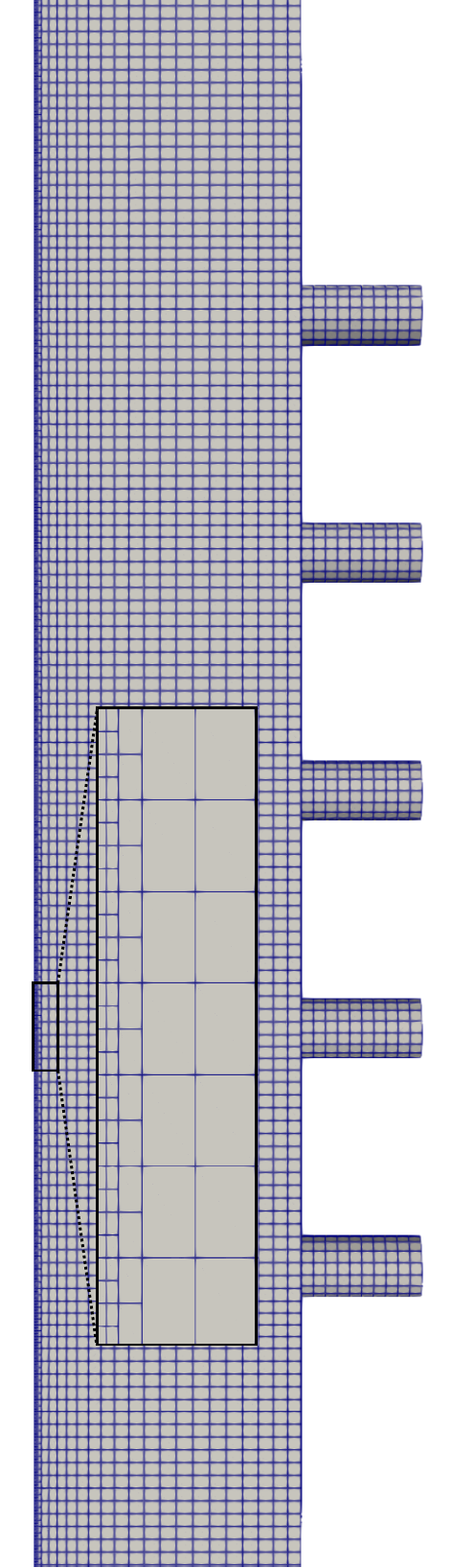}
  \caption{Side view of the coarsest initial 3D mesh (228,000 elements) used for the mesh convergence analysis. The zoomed sub-image illustrates the mesh refinement along the top boundary.}
  \label{fig:mesh}
\end{figure}

We perform a mesh convergence analysis on the arrangement with the highest inlet Reynolds number and maximal cross-flow of the five by one active cooling system, as this is the case we expect to exhibit the greatest flow complexity. This arrangement corresponds to the simulation where jets 0, 1, 2, and 3 are set as inlets and where jet 4 is set as an outlet. We consider three mesh density for the mesh convergence analysis: 228,000 cells, 1.2 million cells, and 3.0 million cells. The mesh with 228,000 cells is shown in Figure~\ref{fig:mesh}.

\begin{figure}[ht]
  \centering
  \includegraphics[width=1\textwidth]{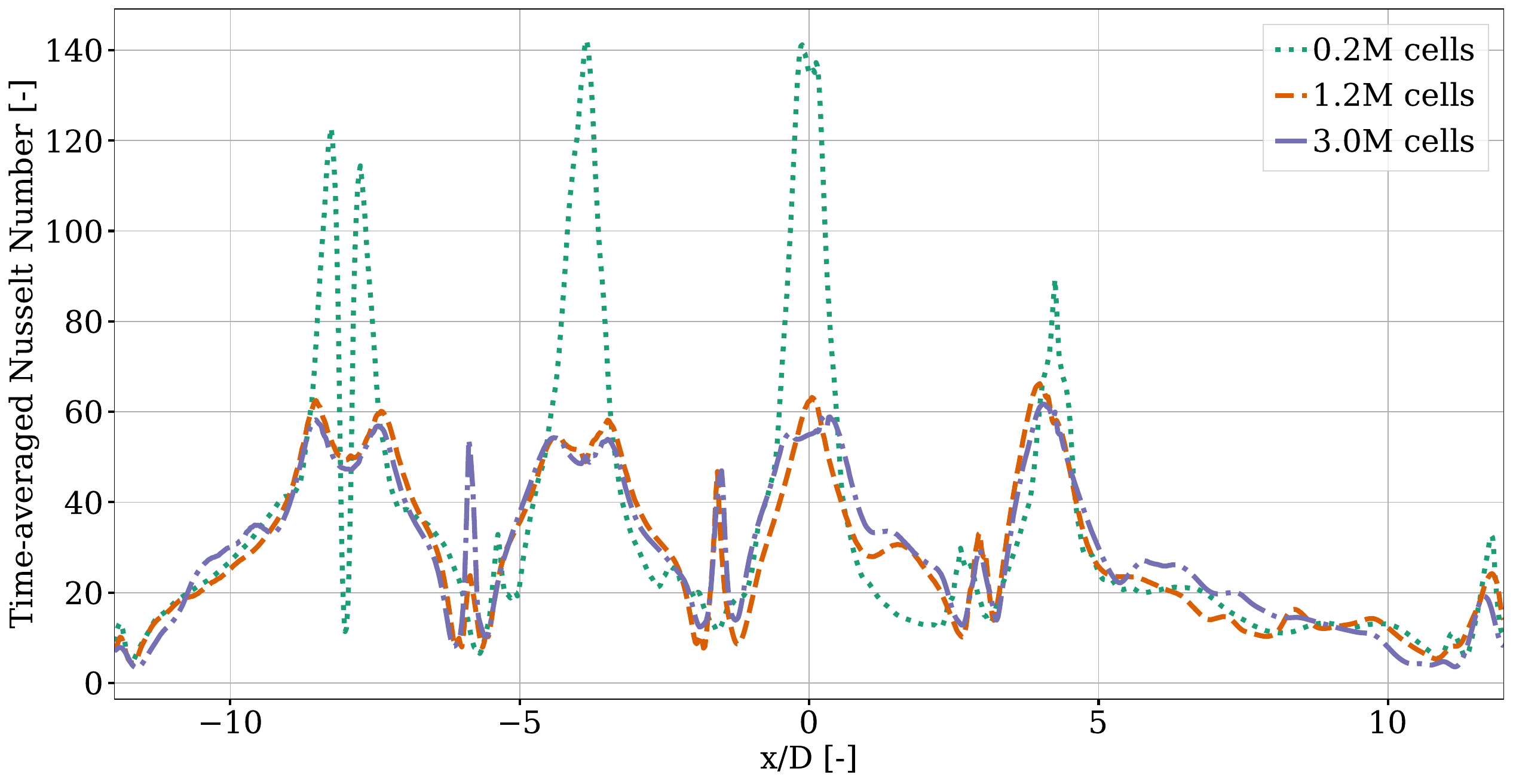}
  \caption{Mesh convergence analysis results where jets 0, 1, 2, and 3 are inlets at a Reynolds number of 2,000 and jet number 4 is set as an outlet. We present three simulation results using 228,000 cells, 1.2M cells and 3.0M cells. The three time-averaged Nusselt number curves are taken along the center of the five by one jet active-cooling system ($y = L/2$). The time average is computed from 40s to 60s. Results show mesh independence with a RMSE of 0.07 for 1.2M cells.}
  \label{fig:mesh_convergence}
\end{figure}

As shown in Figure~\ref{fig:mesh_convergence}, solutions obtained using 1.2M and 3.0M cells exhibit minor differences. The RMSE between the two solutions is 0.07, indicating that satisfactory mesh independence has been reached. Given the geometrical similarities between the five by one and the three by three geometries, we assume the same mesh density is appropriate for the latter.

\subsection{Simulation computational cost}

To train our neural network, we perform all simulations at an inlet Reynolds number below 2,000. This choice is primarily driven by computational cost. Achieving good performance with a neural network typically requires a large amount of training data and generating a sufficiently large dataset becomes more expensive as the Reynolds number increases. To illustrate the impact of the Reynolds number on computational cost, Table \ref{tab:cost} compares the computational cost between a simulation at a Reynolds number of 2,000 and at a Reynolds number of 10,000.

\begin{table}[ht!]
\caption{Cost per simulation at different Reynolds number}
\centering
    \begin{tabular}{@{}cc@{}}
        \toprule
        \textbf{Inlet Reynolds number}    & \textbf{Cost (core-years)} \\ \midrule
        2,000 & 0.03  \\
        10,000 & 0.25 \\ \bottomrule
    \end{tabular}
\label{tab:cost}
\end{table}

At approximately 0.03 core-years per simulation, the total computational cost for a surrogate model training using 100 ILES simulations for a Reynolds number under 2,000 remains reasonable for the five by one configuration (approx. 3 core-years). In contrast, the computational reaches 25 core-years for a surrogate model trained using 100 simulations at a Reynolds number of 10,000. This represents a cost more than eight times greater. To ensure our model remains accessible to users without access to large-scale computing resources, we decided to develop our model using a maximum Reynolds of 2,000. This approach is further motivated by the potential use of the surrogate model for model-based control. In this context, the temperature measurement used in the feedback controller can compensate for the error introduced when scaling predictions to higher Reynolds numbers.

%**********************************************************
% FEM
%**********************************************************

%**********************************************************
% Methodology
%**********************************************************
\section{Neural Network training and hyper-parameter selection}
%***************************
% Neural network
%***************************

\subsection{Surrogate Data Generation}

\begin{figure}[ht!]
    \centering
    \includegraphics[width=1\linewidth]{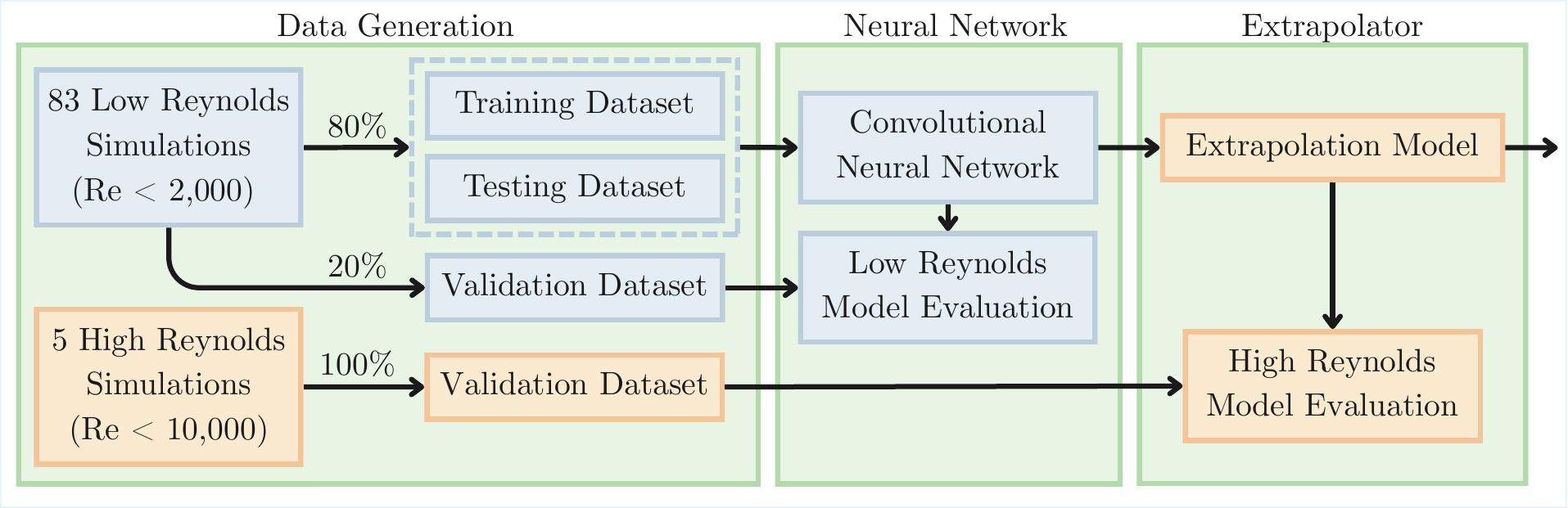}
    \caption{Simulation partitioning for the five by one active cooling system. 83 low Reynolds number simulations (Re < 2,000) are split into training, testing, and validation sets. The training and testing sets are used to train and to select the convolutional neural network's (CNN) hyperparameters. This tuning is done using a k-fold cross validation technique. Performance evaluation of the CNN is done by comparing the predictions to unseen validation simulation results (20\%). The predictions at Re=2,000 of the trained CNN are then extrapolated using correlation based technique to Re=10,000. A second validation dataset is then used to validate the performance of the extrapolation.}
    \label{fig::surrogate_steps}
\end{figure}

As illustrated in Figure \ref{fig::surrogate_steps}, the first step to develop the surrogate model involves the generation of a dataset that is used for training, hyperparameter tuning and validation. To achieve this, we use CFD simulations to compute the Nusselt number profiles on the impingement plate for different arrangements. These simulation results are separated into two groups; simulations that use a maximum inlet Reynolds of 2,000 (Low Reynolds Simulation Data) and simulations using a maximum Reynolds of 10,000 (High Reynolds Simulation Data). 

The low Reynolds number (Re < 2,000) simulation data group depicted in Figure \ref{fig::surrogate_steps} is composed of 83 simulations for the five by one active cooling system and of 100 simulations for the three by three active cooling system. Simulations mainly differ by their arrangement (position of the inlet, position of the outlets and inlet Reynolds number value). All arrangements are generated using the Latin Hypercube Sampling (LHS) method \cite{afzal2017effects}. LHS provides a more uniform and efficient coverage of the parameter space compared to random sampling of injector flow rates and states. These arrangements are the inputs of the neural network and compose a vector of one value per jet (e.g. [-1, -1, 1.0, 0.5, -1]). In this vector, a value of -1 represents an outlet. Any value between 0.0 and 1.0 is the normalized inlet Reynolds number. A value of 0.5 is equivalent to an inlet Reynolds number of 1,000. 

We separate the low Reynolds simulation dataset into a training dataset (80\%) and a validation dataset (20\%) before augmenting the training dataset with the symmetries of the arrangements. For instance, the reflection of the Nusselt number obtained from the configuration [1, -1, -1, -1, -1] is the solution of the [-1, -1, -1, -1, 1] configuration. We tune the hyperparameters of the neural network by applying a k-fold cross validation technique on the training dataset. We use four folds and an 80\%/20\% split (train/test). Finally, we use the validation dataset for the final assessment of the neural network's predictive performance. \cite{cross_validation}

To validate the predictions of the high Reynolds extrapolator model, we use five simulations at Re=10,000 with different configurations (denoted \textit{High Reynolds Simulations}) in Figure \ref{fig::surrogate_steps}. With the exception of the inlet Reynolds number, high Reynolds number simulations use the same simulation parameters as those of the low Reynolds number simulations (see Table \ref{tab:simulation_parameters}).

\subsection{Data Preprocessing}

We apply data preprocessing for two main reasons. First, it allows us to embed additional relevant information into the input data. We provide contextual information by including the spatial coordinates and the inlet or outlet status of each injector. Second, the Nusselt number distribution obtained from simulations has a much higher spatial resolution than what is needed. To address this, we interpolate the simulation results to match the resolution of the infrared thermal camera used in the experimental setup.

To embed more information into the original input vector (e.g. [-1, -1, 1.0, 0.5, -1]), we first replace all the -1 in the input vector and replace them by zeros. This step has been shown to improve performance as the value $-1$ does not have an inherent numerical meaning. By replacing it with zero, we ensure that outlets do not produce unintended signals in the forward pass. To retain information about which jets are acting as inlets or outlets, we introduce a second vector which acts as a filter for jet state. In this vector, we represent inlets by 1 and outlets by $-$1. Finally, we also add a vector to encode normalized distance to the center in the five by one system. For the input [-1, -1, 1, 0.5, -1], we obtain the following 3x5x1 input:
\begin{equation} \label{eq:pre_processed_5x1}
    \left[ \begin{bmatrix} 0 & 0 &1 & 0.5 & 0 \end{bmatrix}, \begin{bmatrix}
-1 & -1 & 1 & 1 & -1 \end{bmatrix}, \begin{bmatrix} -1 & -0.5 & 0 & 0.5 & 1 \end{bmatrix}\right]
\end{equation}

For the three by three active cooling system, the input is transformed into a three by three two dimensional array. This encodes position implicitly in the input as it will go through a series of convolutions during the forward pass of the CNN. The input [-1, -1, -1, 0, 0.25, 0.5, 1, 0.75, 0.5] for the three by three active cooling system would yield the following:
\begin{equation}
    \left[ \begin{bmatrix} 0 & 0 & 0 \\ 0 & 0.25 & 0.5 \\ 1 & 0.75 & 0.5 \end{bmatrix}, \begin{bmatrix} -1 & -1 & -1 \\ 1 & 1 & 1 \\ 1 & 1 & 1 \end{bmatrix}, \begin{bmatrix} -1 & 0 & 1 \\ -1 & 0 & 1 \\ -1 & 0 & 1 \end{bmatrix}, \begin{bmatrix} 1 & 1 & 1 \\ 0 & 0 & 0 \\ -1 & -1 & -1 \end{bmatrix} \right]
\end{equation}

As in \ref{eq:pre_processed_5x1}, the first array contains the normalized inlet  number, the second array contains the state of the injector (inlet is 1 outlet is -1) and the third array is the distance from the center along the x-axis. For the three by three active cooling system, an extra array is appended for the added dimension, as jets are now along the x-axis and the z-axis.

Prior to training, we transform each time-averaged Nusselt number into a two-dimensional Nusselt map with a resolution of 25 by 101 pixels for the five by one model and 75 by 75 pixels for the three by three.

\subsection{CNN Training}

After generating the low Reynolds number cases ($\Reu <$ 2,000), we train a convolutional neural network (CNN) \cite{deconvolution} using the PyTorch library \cite{pytorch}. We use the model to predict the Nusselt number profile on the impinged surface. CNNs are a type of neural network that use convolutional filters (kernels) to learn spatial patterns in data. We opted for a CNN because neighbouring cells in the Nusselt distribution are spatially correlated, and the convolution operations are well suited to capture these local dependencies. A comparison with an ANN is shown in supplementary results (\ref{sup:cnn_vs_ann}).

\begin{figure}[ht!]
    \centering
    \includegraphics[width=1\linewidth]{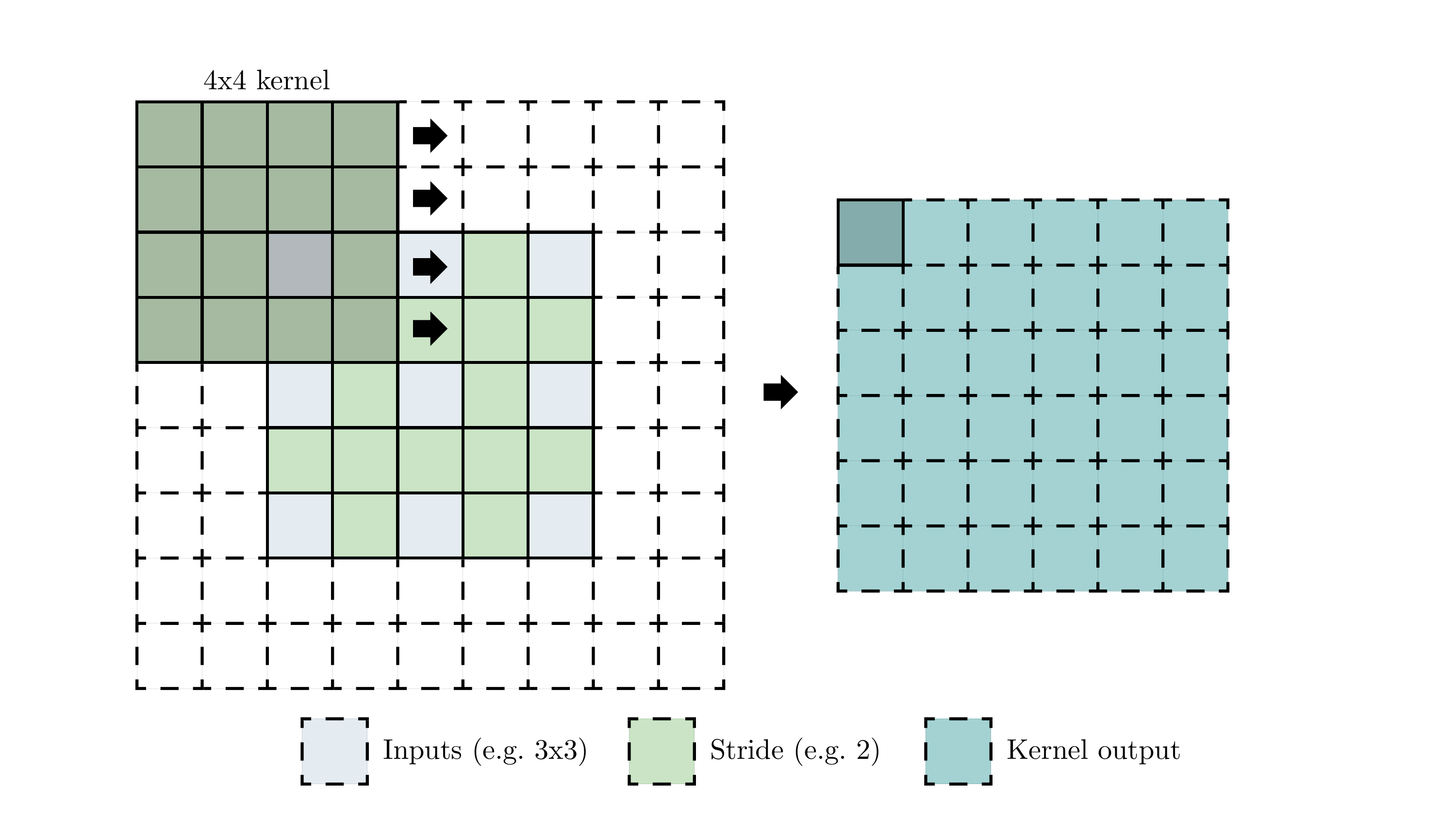}
    \caption{Illustration of a transposed convolution operation with a 4 by 4 kernel and a stride of 2. The input feature map (left) is expanded, and the kernel is applied in a sliding-window fashion to produce a larger output feature map (right). The powdered blue highlights the input values of a single channel, while the cyan is the resulting output of the kernel operation. This operation effectively doubles the size of the input.}
    \label{fig:transposed_convolution}
\end{figure}

\begin{figure}[ht!]
    \centering
    \includegraphics[width=1\linewidth]{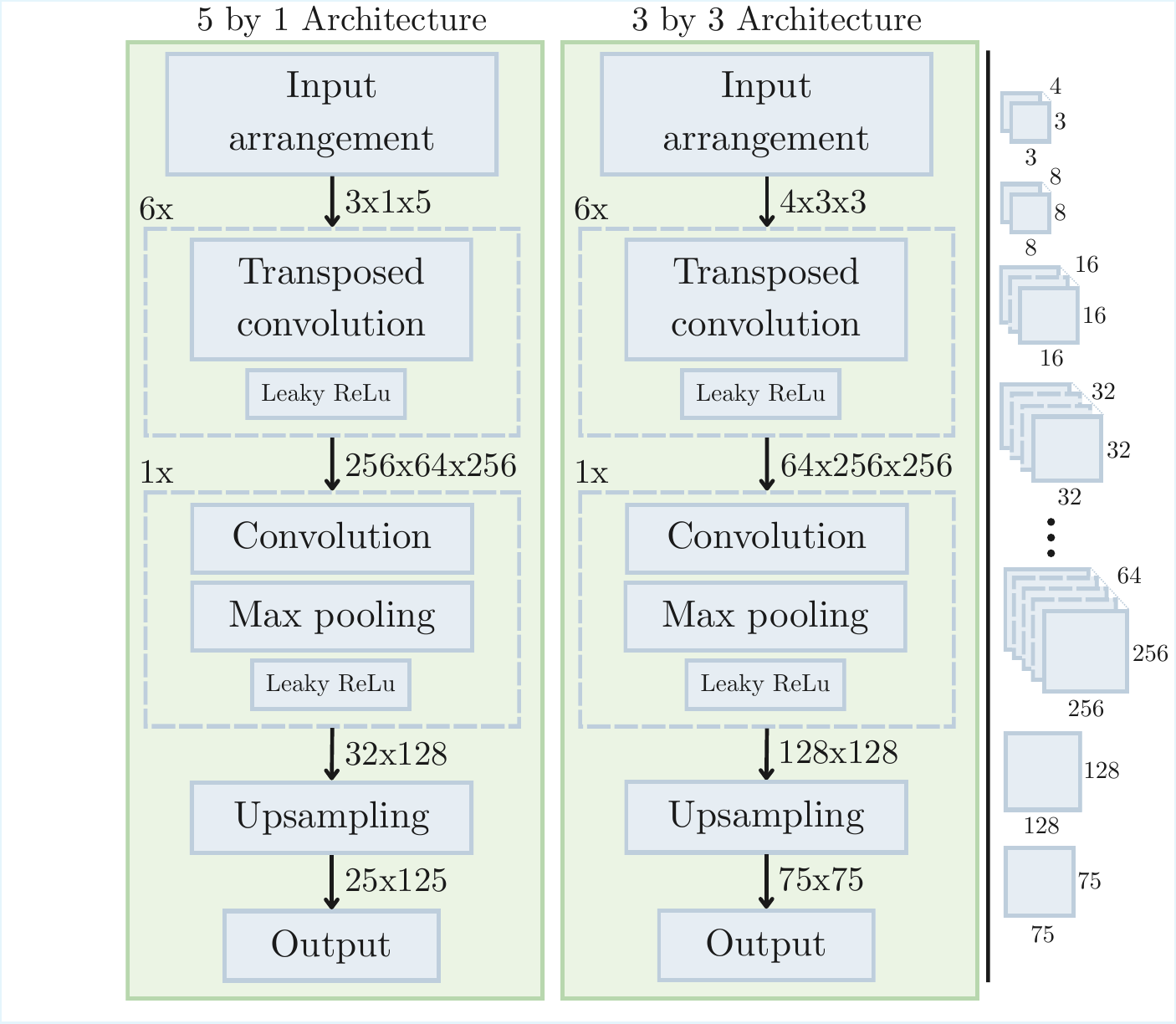}
    \caption{Architecture of the convolutional neural network used for both the five by one and three by three surrogate models. The network architecture consists of a series of upscaling layers of the input jet configuration followed by a convolution layer to reconstruct the Nusselt number distribution (Output) at the target resolution. The input arrangement for the three by three arrangement would start with dimensions 4x3x3. It contains a 3 by 3 array of the normalized Reynolds number, a 3 by 3 array containing the state of the inlets, and two 3 by 3 arrays containing the normalized position along the x-axis and the y-axis. Throughout each layer, array sizes and number of channels are doubled until they reach the maximal image dimensions or the maximal number of channels. Leaky-ReLu activation functions are applied after the transposed convolutions or the convolutions. An upsampling operation is applied at the end to match the output dimensions.}
    \label{fig:architecture}
\end{figure}

Through a series of transposed convolution layers, such as the one illustrated in Figure \ref{fig:transposed_convolution}, we aim to progressively increase the low resolution input to reconstruct a Nusselt number map with a greater amount of pixel than the final output. Specifically, we scale the initial 4x3x3 (normalized Reynolds number, inlet/outlet filter, normalized position along x-axis and normalized position along z-axis) array of the three by three active cooling system into a 64x256x256 array using these operations. We then go through a regular convolution and pooling operation to scale it back to the 75x75 array of the prediction. This process and the complete architectures of the two five by one and three by three CNNs are illustrated in Figure \ref{fig:architecture}. Similar Deep CNN architectures have used similar architectures to enhance noisy image quality \cite{tian2020deep}.
For instance, \citet{mao2016image} employed an architecture that first used convolutions to reduce the spatial dimensions of the input images before restoring their size through transposed convolutions. As our output is already in reduced form, we omit the initial downscaling step and apply only transposed convolutions. 

%The architecture of the convolutional neural network used for both the five by one and three by three surrogate models. The network consists of a series of upscaling layers that expand the input jet configuration, followed by convolutional layers that reconstruct the Nusselt number distribution at the target resolution. Batch normalization and activation functions are applied after each convolution to improve training stability and non-linear feature extraction. The network is trained using mean squared error loss between predicted and CFD-generated Nusselt distributions, allowing it to capture both global and local heat transfer patterns efficiently.

In Figure \ref{fig:architecture}, each of the transposed convolutions uses a kernel size of 4, a stride of 2 and a padding of 1. This combination of parameters ensures that the resulting image dimensions are doubled at every step (e.g. 3x3 $\rightarrow$ 6x6). Between each of these operations, we also double the number of channels, which is equivalent to increasing the number of images on which we apply the kernel operation. Channels are doubled until it reaches the maximum number of channels specified for the architecture. For the five by one model it corresponds to 256 while it corresponds to 64 for the three by three model. 

Convolution and pooling stages are then applied to recover the desired output dimensions. The size of the image is reduced by half, and the total number of channels is reduced to one for the final prediction. The convolution uses a kernel size of 3, a stride of 1 and a padding of 1. We use a pooling kernel with a size of 2 and a stride of 2. It is also important to note that after every transposed convolution or convolution, we apply a non-linear activation function (Leaky ReLu) to the output.

\subsection{Hyperparameter Tuning}

Several hyperparameters shown in Table \ref{tab:nn_hyper} were selected using a k-fold cross-validation method. For instance, the maximum number of channels, the maximum image size during upscaling (before the convolution layers), the dropout rate, and the number of training epochs were all chosen to minimize the RMSE in the test dataset. A grid search was performed to identify the optimal values within the ranges listed in Table \ref{tab:nn_hyper}. Among them, the maximum number of channels had the greatest impact on model performance. Although increasing this parameter generally improved predictive accuracy, no significant gains were observed beyond 256 channels for the five by one model and 64 channels for the three by three.

\begin{table}[ht!]
\caption{Hyperparameters of the neural-network.}
\centering
\begin{tabular}{@{}lccc@{}}
\toprule
\textbf{Parameter}    & \textbf{five by one} & \textbf{three by three} & \textbf{Grid Search} \\ \midrule
 Maximal number of channels & 256 & 64  & [32 - 1024]            \\
 Maximal image dimensions & 64x256 & 256x256  & [128 - 512]             \\
Number of epochs & 500 & 1000 & [0 - 1000]\\
Learning rate  & 0.001 & 0.001 &  [0.005 - 0.0001]       \\
Activation function  & Leaky ReLu  & -  \\
Optimizer   & AdamW & -         \\
Measure of error  & RMSE & - \\ \bottomrule
\end{tabular}
\label{tab:nn_hyper}
\end{table}

\section{Extrapolator}

Generating a dataset from CFD simulations at inlet Reynolds numbers up to 2,000 is computationally costly and can significantly limit the number of simulations due to limited computational resources. Hence, this strategy could appear inadequate when attempting to model impinging-jet arrays at higher Reynolds numbers. Alternatively, \citet{martin1977heat} showed that the Nusselt number is proportional to $\Reu^{\alpha}$ where $\alpha$ is approximately 0.574. Nusselt number predictions generated by the neural network at low Reynolds number could therefore provide valuable information for estimating the Nusselt number at higher Reynolds numbers. Equation \ref{eq:extrapolator} uses this property to estimate the Nusselt number at a higher Reynolds number, $\Reu_{\text{pred}}$:

\begin{equation} \label{eq:extrapolator}
    \text{Nu}(\Reu_{\text{pred}}) = \text{Nu}(2000) \left(\frac{\Reu_{\text{pred}}}{2000}\right)^{0.574}
\end{equation}

%**********************************************************
% Results and Discussions
%**********************************************************
\section{Results and Discussion}
%***************************
% Results and Discussions
%***************************

This discussion focuses on the predictions generated by the five by one and the three by three surrogate models. We begin by presenting a selection of predictions on the validation dataset of the outputs of the models. This is followed by a more detailed analysis using relevant error metrics. Finally, we assess the performance of the surrogate models in extrapolation scenarios ($\textrm{Re} < 10,000$) and present results from experimental validation.

\begin{figure}[ht!]
    \centering
    \begin{subfigure}[b]{0.8\linewidth}
        \centering
    \includegraphics[width=\linewidth]{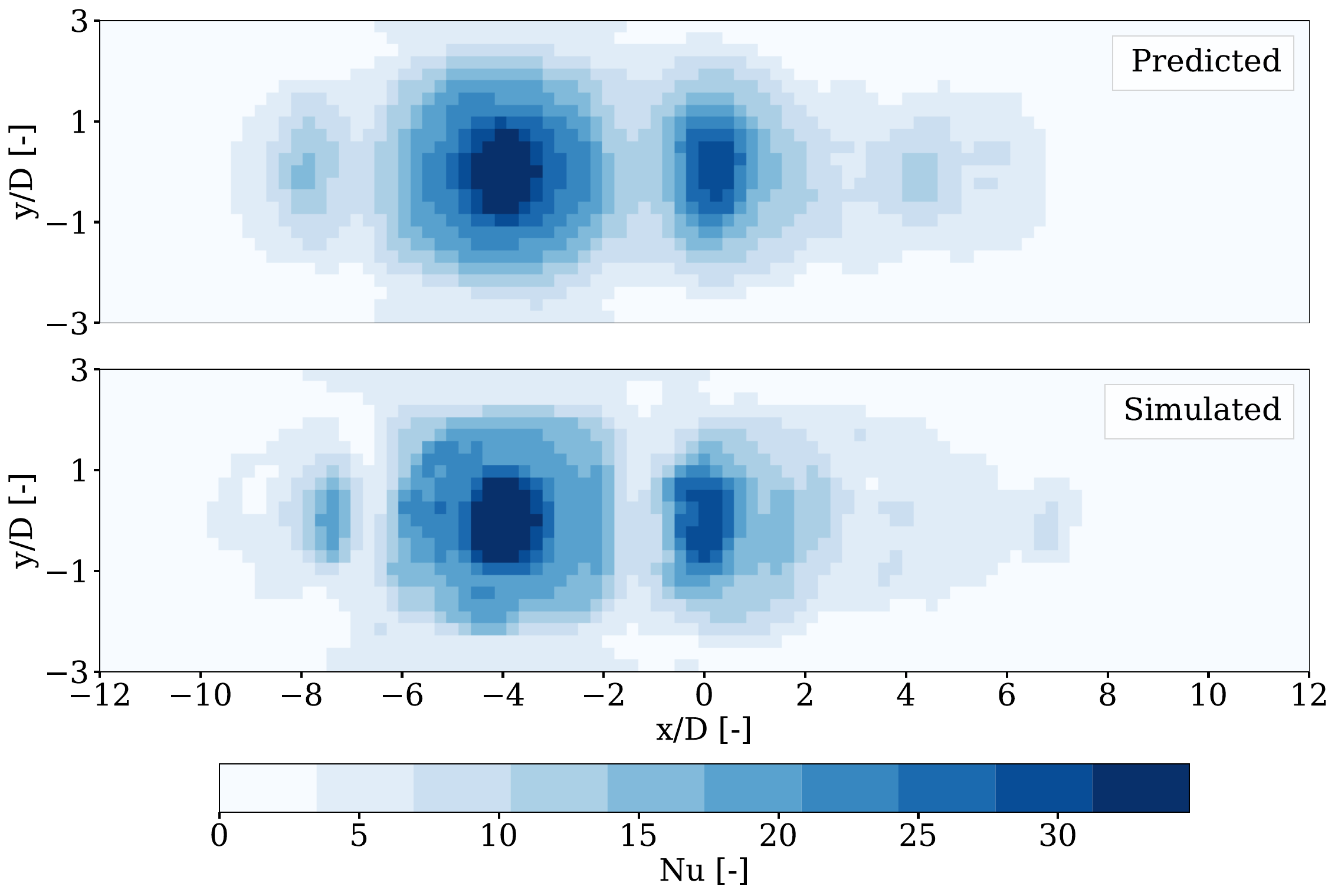} \label{fig:first_config}
        \caption{Surrogate and simulation Nusselt number distribution for the five by one system with configuration [0.16, 0.93, 0.55, 0.17, -1] and Reynolds number below 2,000.}
        \label{fig:5x1-case1}
    \end{subfigure}
    
    \vspace{1em}

    \begin{subfigure}[b]{0.8\linewidth}
        \centering
        \includegraphics[width=\linewidth]{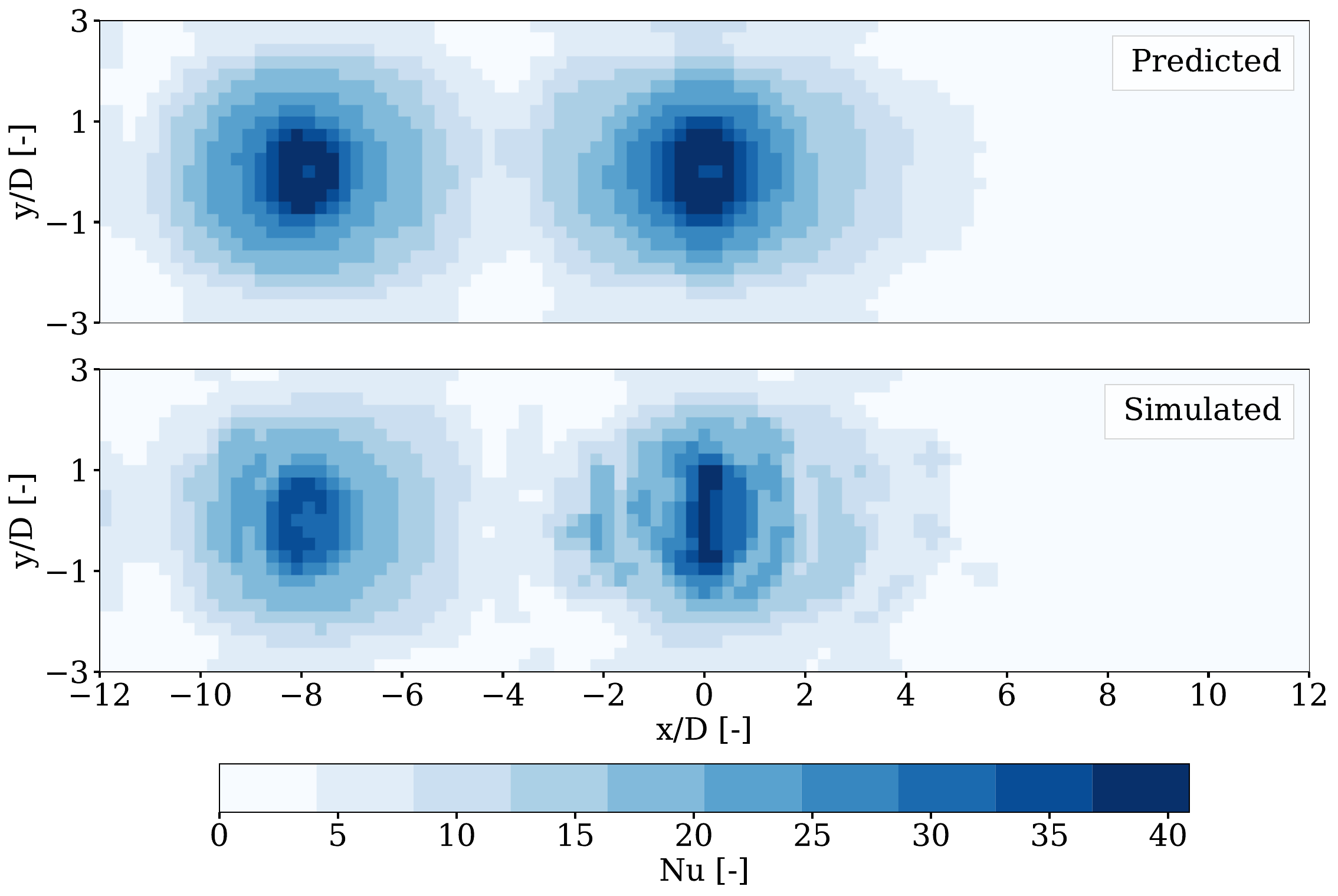}
        \caption{Surrogate and simulation Nusselt number distribution for the five by one system with configuration [0.88, -1, 0.93, -1, -1] and Reynolds number below 2,000.}
        \label{fig:5x1-case2}
    \end{subfigure}
    
    \caption{Comparison between surrogate predictions and simulation results of the Nusselt number distribution for two different configurations in the five by one active cooling system. The Nusselt maps are shown along the impingement surface using dimensionless distances $x/D$ and $y/D$}
    \label{fig:5x1-2D-prediction}
\end{figure}

\begin{figure}[ht!]
    \centering
    \begin{subfigure}[b]{0.8\linewidth}
        \centering
        \includegraphics[width=\linewidth]{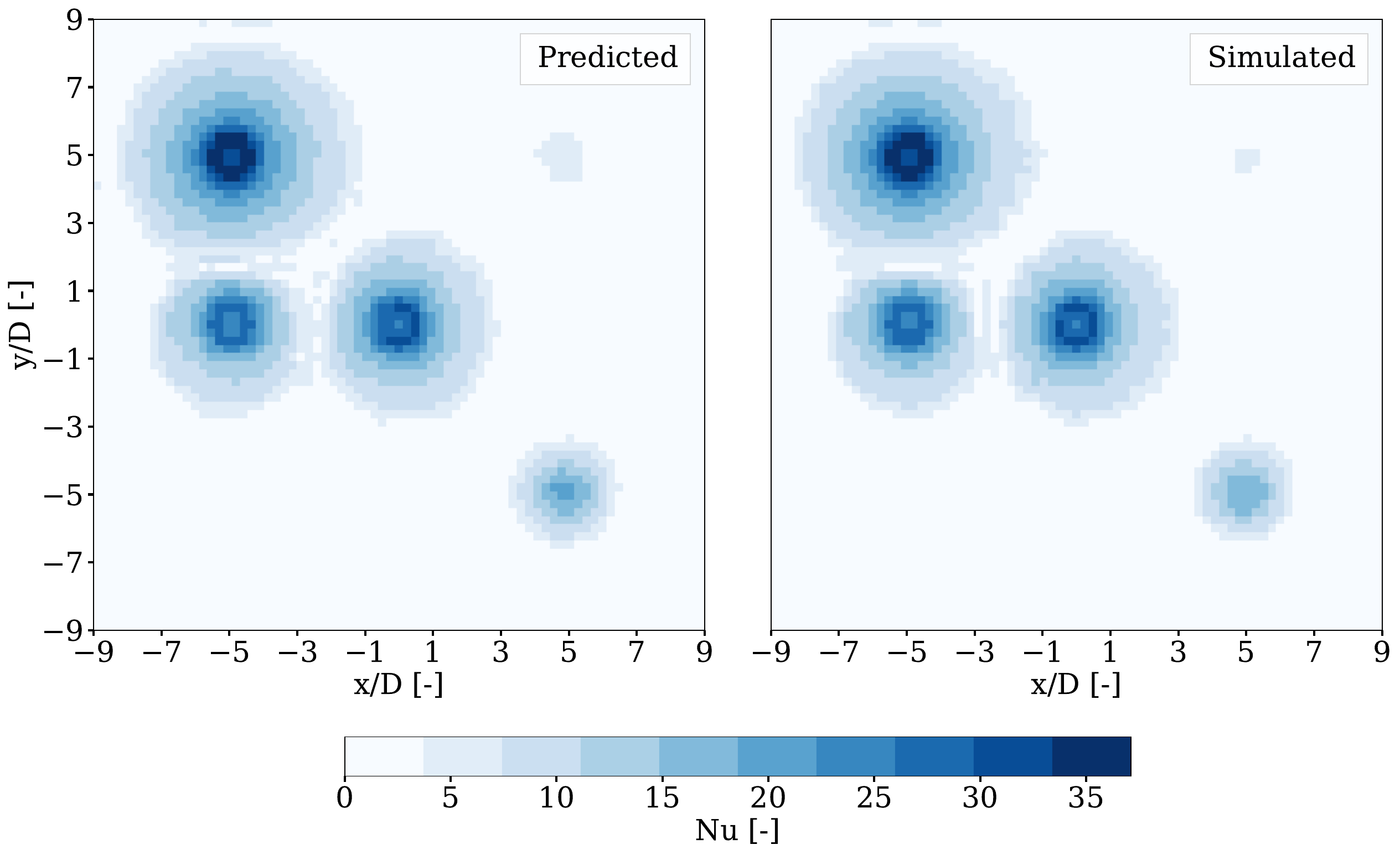}
        \caption{Surrogate and simulation Nusselt number distribution for the three by three system with configuration [0.79, -1, 0.08, 0.45, 0.52, -1, 0.05, -1, 0.16].}
        \label{fig:3by3-case1}
    \end{subfigure}
    
    \vspace{1em}

    \begin{subfigure}[b]{0.8\linewidth}
        \centering
        \includegraphics[width=\linewidth]{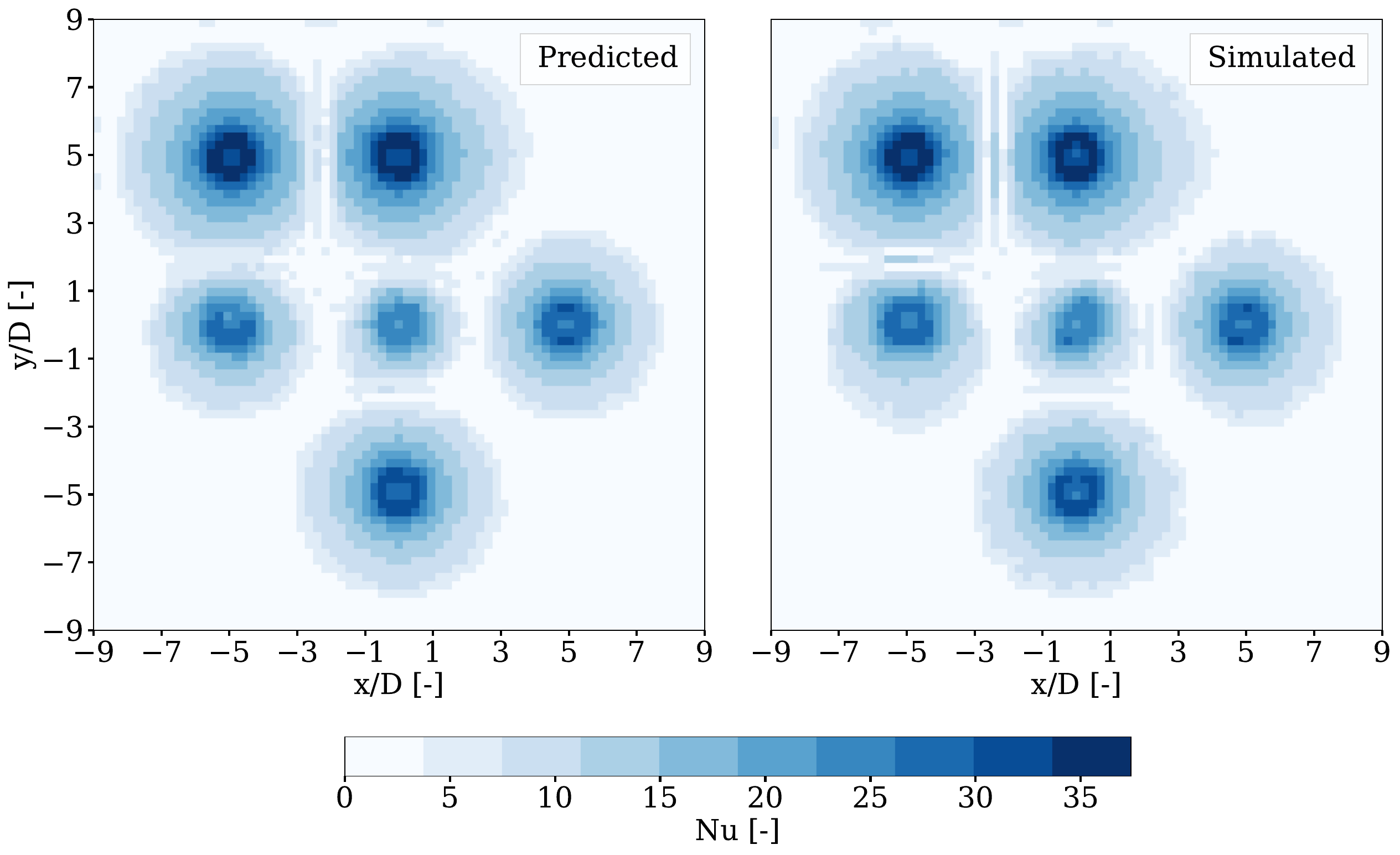}
        \caption{Surrogate and simulation Nusselt number distribution for the three by three system with configuration [0.82, 0.79, -1, 0.48, 0.34, 0.52, -1, 0.61, -1].}
        \label{fig:3by3-case2}
    \end{subfigure}
    
    \caption{Comparison between surrogate predictions and simulation results of the Nusselt number distribution for two different configurations in the three by three active cooling system. The Nusselt maps are shown along the impingement surface using dimensionless distances $x/D$ and $y/D$, where $D$ is the inlet diameter.}
    \label{fig:3by3-2D-prediction}
\end{figure}

Predictions for both the five by one and three by three surrogate models are shown in Figures \ref{fig:5x1-2D-prediction} to \ref{fig:3by3-2D-prediction}. These results are taken from the validation dataset, which was not used during training. As such, they provide an unbiased evaluation of the performance of the models, in contrast to results obtained on the training data where the models were optimized. For the five by one model, we select two arrangements of inlet, outlet, and flow rates that highlight the neural network's ability to capture important features of the Nusselt number distribution. The first configuration, shown in Figure \ref{fig:5x1-case1}, corresponds to the setup [0.16, 0.93, 0.55, 0.17, -1] which includes four inlets and a single outlet. The second configuration, illustrated in Figure \ref{fig:5x1-case2}, features two inlets and three outlets. It is described by the arrangement [0.88, -1, 0.93, -1, -1]. We ensured that the selected configurations are representative of the validation dataset, as the associated prediction errors were not among the highest or lowest observed values.

In Figure \ref{fig:5x1-2D-prediction}, we observe that the surrogate model captures the main features of the heat transfer distribution, the amplitude of these features and some of the interactions between the jets. Due to the forced outflow (single outlet) on the right side in Figure \ref{fig:5x1-case1}, the configuration [0.16, 0.93, 0.55, 0.17, -1] induces a cross-flow that alters the Nusselt number profile of the central jet and that is captured by the surrogate (approx. $x/D=4$ and $y/D=0$). This case highlights the complexity of these flows and demonstrates the utility of deep learning techniques in modeling the active cooling system.

While the surrogate model captures the general structure of the Nusselt number, some high frequency fluctuations (approx. $x/D=-2$ and $y/D=0$) present in the simulated Nusselt number in Figure \ref{fig:5x1-case2} are not well-captured by the predictions of the five by one model. At the end of the current section, we show that these high-frequency fluctuations have a minimal impact on the predicted average Nusselt number, suggesting that the surrogate model remains reliable for capturing the dominant heat transfer characteristics despite local discrepancies in the five by one model. One more limitation of the surrogate model can be observed in Figure \ref{fig:5x1-case1}. In this simulation, the peak associated with the fourth inlet from the left appears to be absent, likely due to jet-jet interactions. On the contrary, the surrogate model predicts a small peak in that region. This behaviour appears to be mitigated by increasing the size of the training dataset. This is showcased in the 3x3 surrogate model where such interactions are captured more adequately. The three by three model was trained on more training data as the geometry contains more axes of symmetry (8 for 3D vs 2 for 2D).

Predictions made by the 3x3 model are shown in Figure \ref{fig:3by3-2D-prediction}. We chose two configurations, [0.79, -1, 0.08, 0.45, 0.52, -1, 0.05, -1, 0.16] and [0.82, 0.79, -1, 0.48, 0.34, 0.52, -1, 0.61, -1]. Each of them exhibits jet-jet interaction and are representative of our validation dataset.

A summary of important performance metrics for the surrogate models are listed in Table \ref{tab:performance}. The normalized mean average error (NMAE) is defined as:

\begin{equation} \label{eq:nmae}
    \text{NMAE} = \frac{1}{n} \sum_i^n \frac{|y_{sim}-y_{pred}|}{\max(y_{sim})}
\end{equation}

The average maximum error corresponds to the mean of the largest pixel deviations observed in each prediction from the validation dataset. The Average top 10\% error is calculated using the mean error of the average 10\% of the pixels displaying the largest error in each predictions of the validation dataset. 

\begin{table}[ht!]
\caption{Surrogate performance metrics calculated using the simulated and predicted Nusselt number on the impinged surface. All metrics are calculated using results from the validation data set.}
\centering
\begin{tabular}{@{}lcc@{}}
\toprule
\textbf{Perfomance Metrics}    & \textbf{five by one} & \textbf{three by three} \\ \midrule
Validation RMSE [-] & 0.64 & 0.24 \\
Validation NMAE [-] & 0.016 & 0.0057 \\
Average maximum error [-] & 4.8 & 3.6 \\
Average top 10\% error [-] & 2.5  & 1.2  \\ \bottomrule
\end{tabular}
\label{tab:performance}
\end{table}

As shown in Table \ref{tab:performance}, the three by three surrogate model outperforms the five by one in all metrics. This is mainly caused by the larger training dataset given the amount of symmetries in the three by three system. The average error and the average top 10\% error provide an estimate of the largest errors that the model is likely to produce. These errors are generally the highest near the impinging point of the jets.

\begin{figure}[ht!]
    \centering
    \includegraphics[width=0.8\linewidth]{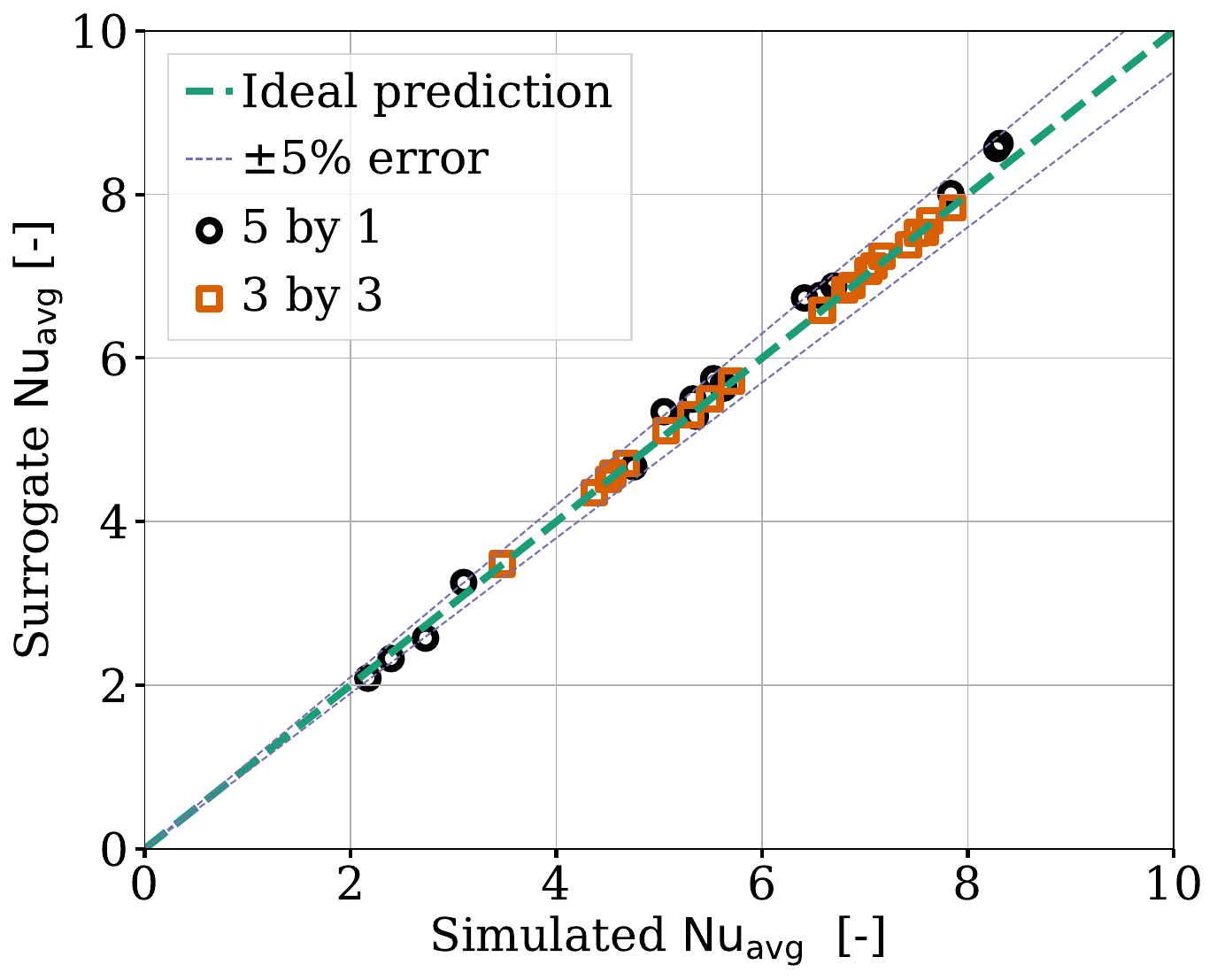}
    \caption{Parity plot of the average Nusselt number. Nu$_\textrm{avg}$ corresponds to the spatial average Nusselt on the top of the cavity (green dashed line in Figure \ref{fig:cooling_geometries}). Every point corresponds to a single simulation of the validation dataset presented in figure \ref{fig::surrogate_steps}. An ideal prediction would stand on the green dashed line.}
    \label{fig:parity_plot}
\end{figure}

We want the model to accurately predict the total Nusselt number over the entire surface. Small local variations in the Nusselt distribution, such as peak shifting, can sometimes artificially increase the local error. However, it is also important to assess whether these discrepancies introduce any bias in the global average Nusselt number. Hence, Figure \ref{fig:parity_plot} shows a parity plot for the differences in the average heat flux and the simulation results of the validation dataset for both the 5x1 and 3x3 model. For the five by one model, the model introduces a small bias. 11 predictions (black circle) are above the ideal prediction line while only 5 are below. On the other hand, the three by three model introduces no significant bias. 

Only two predictions fall outside of the $\pm$5\% average error margin for the five by one, while none of the three by three predictions fall outside the $\pm$5\% line. In fact, the maximum error for the three by three surrogate is 1.1\%. This is significantly lower than the maximum 5.8\% error of the five by one surrogate model. The average error for the Nusselt number average is 0.2\% for the three by three and 1.4\% for the five by one. Although the surrogate models may introduce local prediction errors, this shows that the global Nusselt number is well conserved.

\subsection{Extrapolation performance at $\mathrm{Re}=10,000$}

\begin{figure}[ht!]
    \centering
    \begin{subfigure}[b]{0.8\linewidth}
        \centering
        \includegraphics[width=\linewidth]{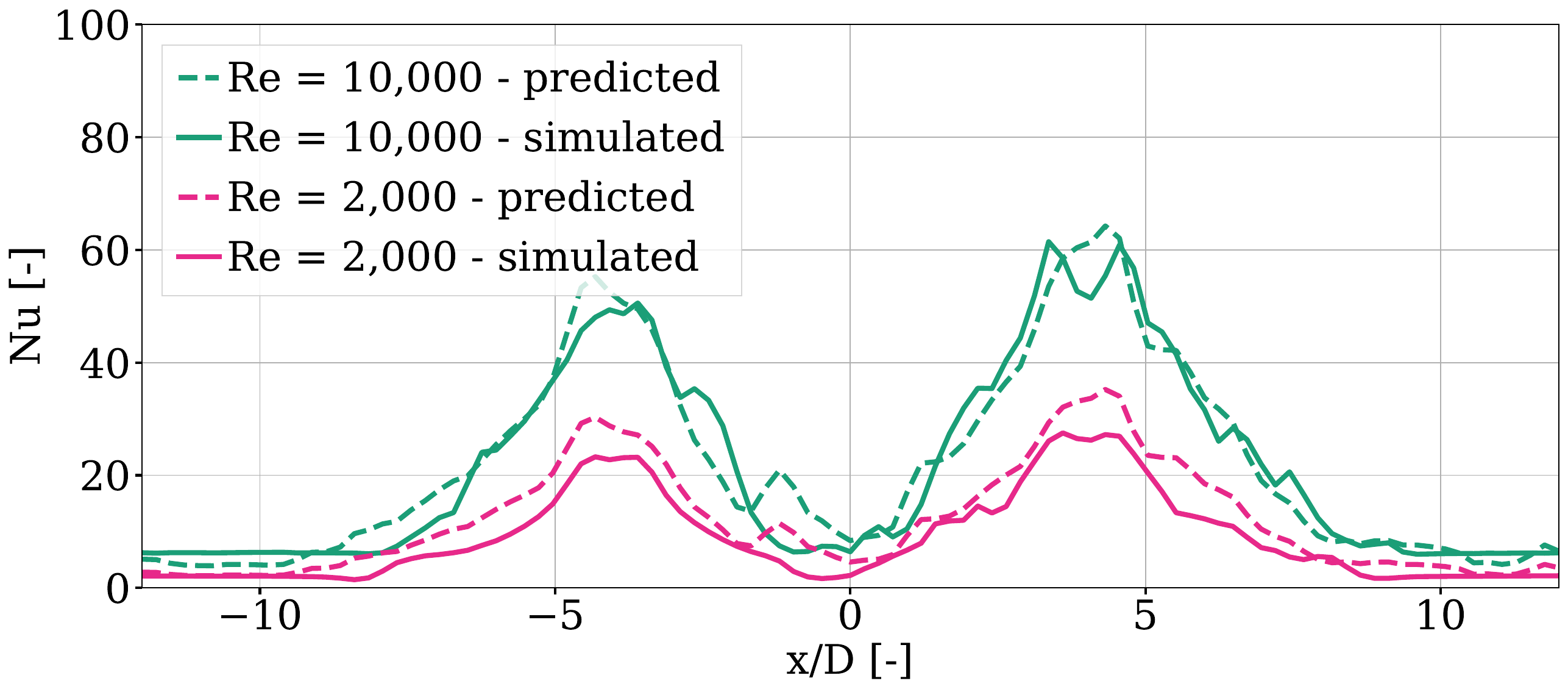}
        \caption{Simulated and scaled predictions for Reynolds number under 2,000 and 10,000 for the configuration [-1, 0.37, 0.06, 0.57, -1].}
        \label{fig:extrapolation_0}
    \end{subfigure}
    
    \vspace{1em}

    \begin{subfigure}[b]{0.8\linewidth}
        \centering
        \includegraphics[width=\linewidth]{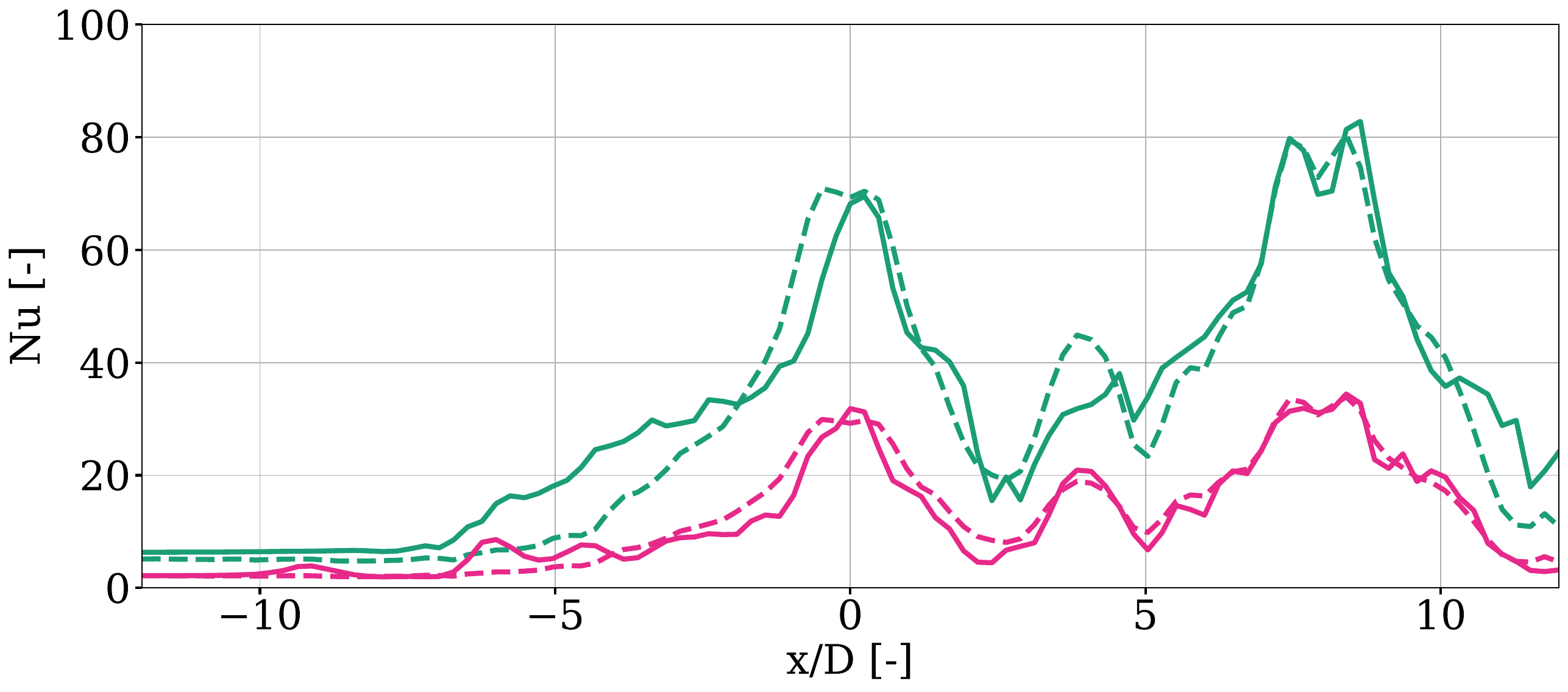}
        \caption{Simulated and scaled predictions for Reynolds number under 2,000 and 10,000 for the configuration [-1, -1, 0.63, 0.21, 0.90].}
        \label{fig:extrapolation_1}
    \end{subfigure}

    \vspace{1em}

    \begin{subfigure}[b]{0.8\linewidth}
        \centering
        \includegraphics[width=\linewidth]{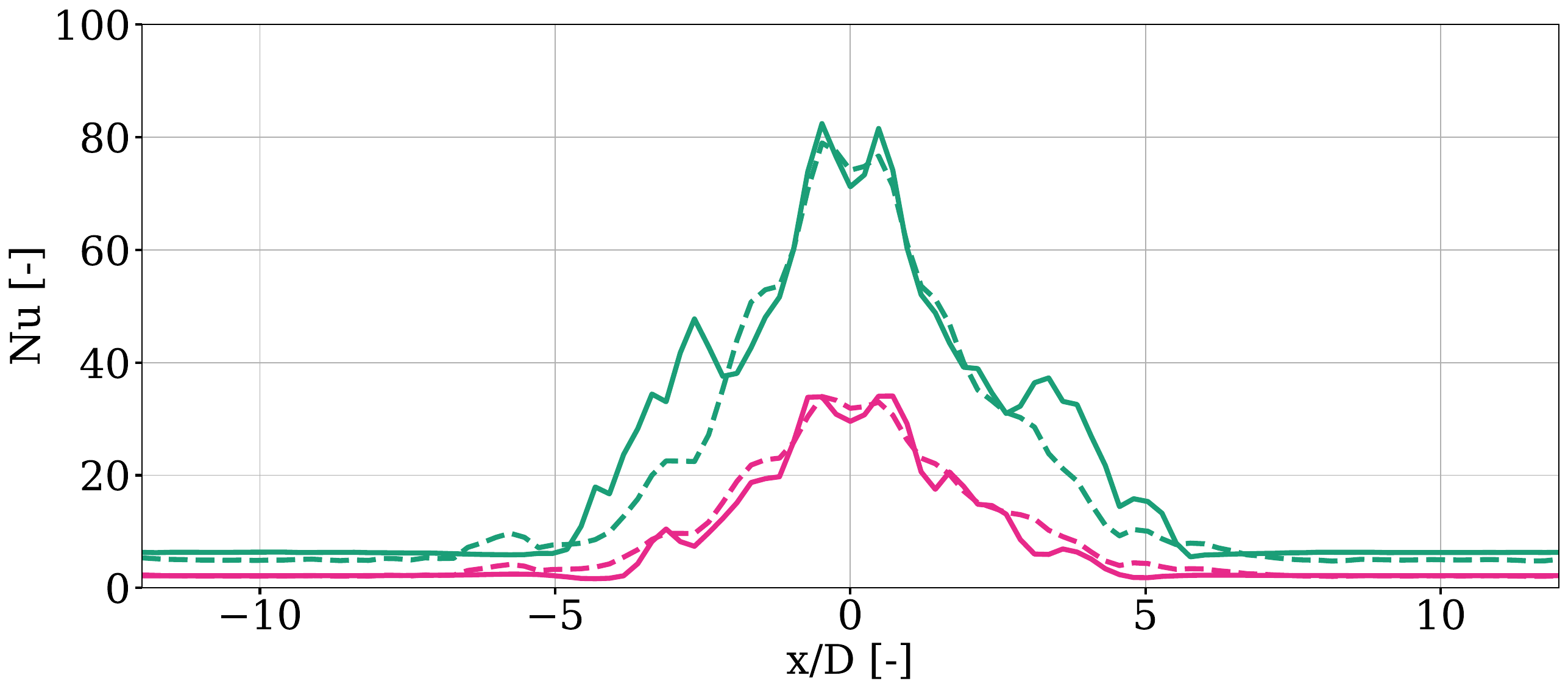}
        \caption{Simulated and scaled predictions for Reynolds number under 2,000 and 10,000 for the configuration [-1, 0.05, 0.87, -1, -1].}
        \label{fig:extrapolation_2}
    \end{subfigure}
    
    \caption{Comparison between surrogate predictions and simulation results before and after scaling of the Nusselt number distribution. Results are taken on a line along the middle of the impingement plate ($y=L/2$). The dashed result at Re=10,000 (green) is obtained with Eq. \ref{eq:extrapolator}. Nu(2000) in Eq. \ref{eq:extrapolator} corresponds to the dashed result at Re=2,000 (pink).}
    \label{fig:extrapolation}
\end{figure}

Figure \ref{fig:extrapolation} shows the difference between the simulated results and the neural network predictions for a five by one configuration on a line along the impingement surface. We compare the predictions of three configurations at a Reynolds number less than 2,000. We then extrapolate the prediction obtained at a Reynolds number less than 2,000 using Equation \eqref{eq:extrapolator} to predict the same configuration at a Reynolds number less than 10,000. The scaled solution is then compared to a simulation executed using the same configuration at a Reynolds number less than 10,000.

In Figure \ref{fig:extrapolation}, the main features of the Nusselt number distribution are still captured after extrapolation. However, the NMAE is slightly higher, reaching a value of 0.11 compared to the 0.016 obtained before the scaling. The RMSE is also higher, at 8.4. Although these error values are greater than before, we expect that the temperature measurement used in the feedback controller can compensate for the difference between the actual system behaviour and the surrogate predictions.

% However, the main limitation of this methodology is that any error present in the prediction at $\Reu=$ 2,000, such as the slight shift of the peak, is directly propagated into the extrapolated results. Additionally, some features such as the one on the left of the third jet in Figure \ref{fig:extrapolation_1} only appear once the Reynolds reaches a certain threshold. A similar observation is made in Figure \ref{fig:extrapolation_2}, where the characteristic double peak commonly seen at high Reynolds numbers cannot be captured by the model when evaluated at low Reynolds numbers. These limitations are reflected in the NMAE, which is slightly higher, reaching a value of 0.11 compared to the 0.016 obtained before the scaling. The RMSE is also higher, at 8.4. Although these values are greater than before, we expect that the temperature measurement used in the feedback controller can compensate for the difference between the actual system behaviour and the surrogate predictions.

\section{Experimental Validation}
% à ajouter, temps adimensionels des expériences. Justifications pour négliger le rayonnement et la convection naturelle . Ajouter le nombre de mesures par secondes 

To validate the surrogate model predictions, we use an experimental setup which is identical to the 5 by 1 active cooling device in Figure \ref{fig:5x1_geometry}. As a validation test case, we use the configuration [1, 1, -1, -1, -1] at a Reynolds number of 10,000. We apply an initially unknown heat flux on the top of the plate using a heat gun and measure the temperature using a thermal camera. The steel plate is painted with a black, high-emissivity coating (Tremclad® High Heat Enamel, black, very flat) to minimize the need for emissivity calibration of the thermal camera.

\begin{figure}[ht]
  \centering
  \includegraphics[width=1\textwidth]{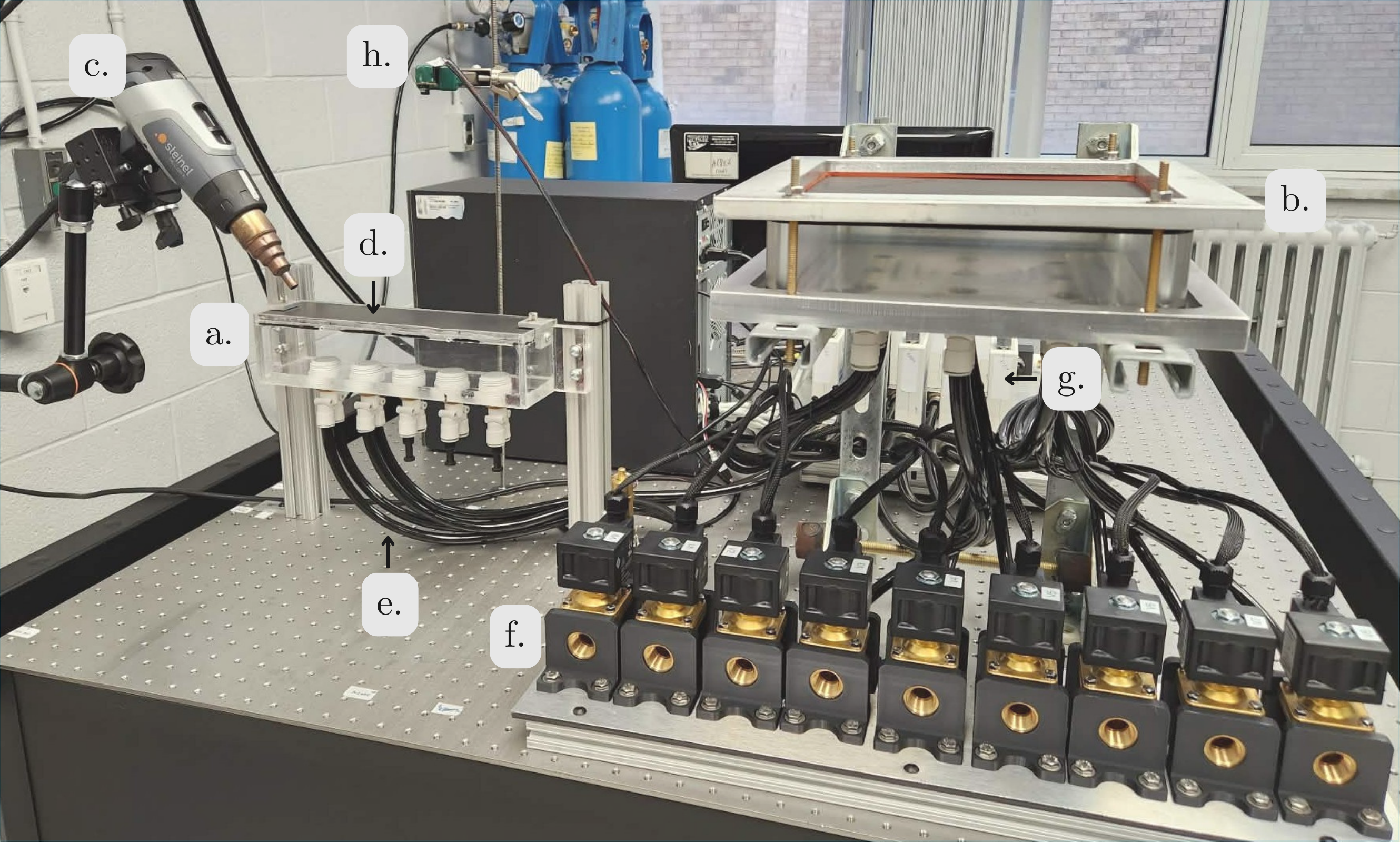}
  \caption{Experimental setup of the 5 by 1 (a.) and 3 by 3 (b.) active cooling systems. During experimental validation, we use a heat gun (c.) to heat the black surface (d.) of the experimental setup on the left (a.). We cool the surface using air jets fed through black tubes (e.). The state and flow rate of these jets are controlled using solenoid valves (f.) and mass flow controllers (g.). Two inlets are plugged in the five by one setup to allow for two inlets at a Reynolds number of 10,000. A thermal camera (h.) is set above the jet array to measure the temperature. The entire experimental setup is described in \cite{oliveira2025active}.}
  \label{fig:validation}
\end{figure}

We replicate the experiment in simulation by solving the enthalpy conservation equation \eqref{eq::enthalpy_conservation} in the solid impingement plate using a 3D second-order accurate finite difference solver. The bottom boundary is defined by the Nusselt number predicted by the surrogate model, the lateral sides use no-flux conditions, and the top applies a boundary condition and a reconstructed convective coefficient of the heat gun. This coefficient is obtained by solving the time-dependent adjoint problem for the no-jet case ([0, 0, 0, 0, 0]) for 60 seconds, following the methodology described by \cite{razzaghi_adjoint}. When solving the time-dependent adjoint problem, we neglect natural convection on the bottom boundary as the quasi totality of the energy provided by the heat gun is accumulated in the plate due to its thermal inertia.

To obtain the reconstructed temperature shown in Figure \ref{fig:validation}, natural convection is included at the top boundary in addition to the reconstructed convective heat transfer coefficient associated with the heat gun.

In figure \ref{fig:validation}, the reconstructed temperature is on average within 5.8\% of the experimental values after 120 seconds. Simulation uncertainties such as the own surrogate model errors and modeling hypothesis are probably responsible for the steady increase of the error in time. However, we aim to use this surrogate model as a tool for model-based control. Since controllers can compensate for certain modeling errors, we expect the observed difference between the experimental and reconstructed temperatures to be mitigated when implementing closed-loop temperature control algorithms.

\begin{figure}[ht]
  \centering
  \includegraphics[width=1\textwidth]{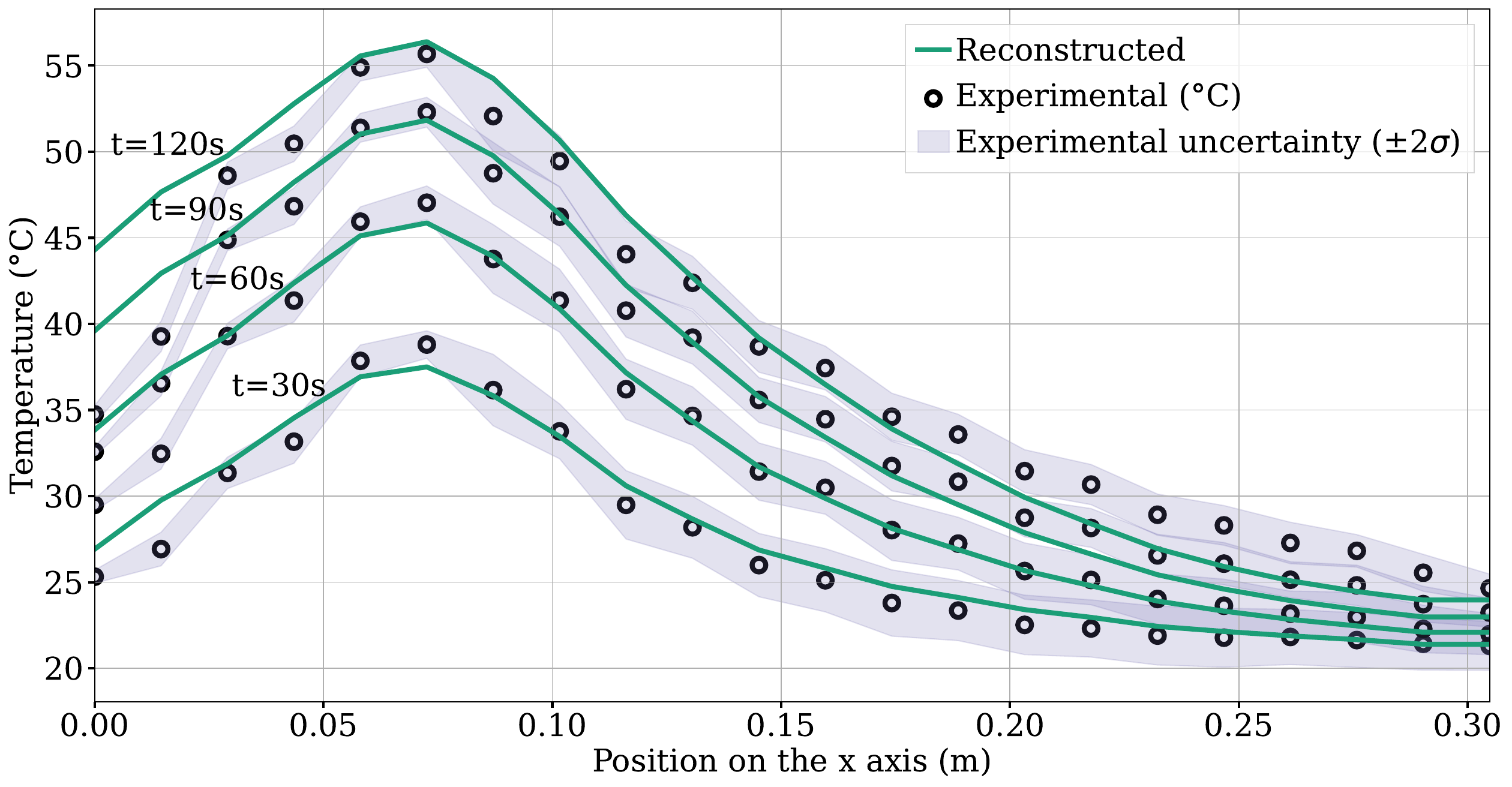}
  \caption{Experimental vs. reconstructed temperature across the x-axis at the center ($y=L/2$) of the active cooling system for the configuration [1, 1, -1, -1, -1] at a Reynolds number of 10,000. The Reconstructed result uses the reconstructed heat flux of the heat gun at the top and the surrogate heat flux at the bottom. The experimental result was recorded using a thermal camera. Experimental uncertainties were calculated by combining the standard deviation of three distinct experiments with the uncertainty obtained from the three reconstruction experiments.}
  \label{fig:validation}
\end{figure}

%**********************************************************
% Conclusion
%**********************************************************
\section{Conclusions}
%***************************
% Conclusion
%***************************

In this work, we developed a surrogate model that predicts the Nusselt number distribution on the impingement surface of two distinct active cooling systems (five by one jets and three by three jets). In these active cooling systems, the flow rates of the jets are independently controlled. The state of the jets, meaning if a single jet is used as an inlet or an outlet, can also be modified. For the surrogate model, we trained a convolutional neural network (CNN) using data generated from computational fluid dynamics (CFD) simulations performed with Lethe, an open-source CFD software based on an implicit Large-Eddy Simulation (ILES) strategy. The simulation dataset for the five by one configuration comprises 83 distinct arrangements, while the three by three dataset includes 100. We executed all the simulations for arrangements of flow rates, inlet and outlets for a Reynolds number under 2,000. We reserved 20\% of the training data for validation and used the remaining 80\% to train and tune the hyperparameters of the model using a 4-fold cross-validation technique.

We showed that this CNN-based surrogate model successfully captures the main features of the Nusselt number distribution on the validation dataset, specifically the location of the heat transfer peaks and the influence of cross-flow between jets. The five by one model achieves a root mean square error (RMSE) of 0.64, while the three by three model performs better with an RMSE of 0.24. We attribute this difference in performance to the amount of training arrangements in the dataset. The three by three model has more symmetries that we use to enhance the training dataset. Despite good performance, limitations remain. We found that the surrogate model can struggle to predict local high frequency variations of the Nusselt number in the five by one configuration. However, these local high frequency variations that are not captured by the surrogate do not appear to significantly influence the average predicted Nusselt number on the plate as only 2 validation cases of the five by one model generate an error higher than 5\%. On average, the error in the average Nusselt number for the five by one is 1.4\% while it is 0.2\% for the three by three active cooling system.

We also tested the ability of our predictions to be extrapolated to a Reynolds number of 10,000. For the extrapolation, we adapted the correlation proposed by \citet{martin1977heat}. We showed that this methodology can introduce some error. In fact, the NMAE for the five by one model goes from 0.016 up to 0.11 during the extrapolation from $\Reu = 2,000$ to $\Reu = 10,000$. Although significant, we aim to use this model in the context of model-based temperature control. We therefore expect this error to be mitigated by the controller.    

In an experimental context, we also found good agreement with the surrogate model predictions. The error is on average within 5.8\% of the experimental values after 120 seconds.

% We conclude that a surrogate model trained on CFD simulations to predict the Nusselt number in impinging jet arrays provides a solid foundation for the development of temperature control strategies. Given its demonstrated performance, future works will focus on using this model in the context of model-based temperature control. This approach is well suited for estimating the Nusselt number in active cooling systems such as the one described by \citet{Lamarre_2023}, where the large number of possible jet arrangements makes the use of correlations impractical.

The main contribution of this work lies in establishing a reproducible framework that uses high-fidelity CFD simulations to train a CNN-based surrogate model that can accurately predict the Nusselt number of a multi-input, multi-output active cooling system up to a Reynolds number of 10,000. Future work will focus on integrating this surrogate model into model-based, closed-loop temperature control strategies. This will allow real-time actuation of individual jet flow rate and configuration in active cooling systems such as the one described by \citet{Lamarre_2023}.

%**********************************************************
% Data avail
%**********************************************************
\section{Data and Code Availability}
The data and code supporting this study are available in a public repository at Zenodo, DOI: 10.5281/zenodo.18157225. \cite{vaillant_2026_18157226}

%**********************************************************
% Acknowledgements
%**********************************************************

\section{Acknowledgments}
%***************************
% Acknowledgments
%***************************

This work was funded by the National Research Council of Canada's (NRC) National Program Office (NPO) through the Advanced Manufacturing program. This research was enabled in part by support provided by Calcul Qu\'ebec (https://www.calculquebec.ca/) and the Digital Research Alliance of Canada (https://alliancecan.ca/en).

\section{Declaration of generative AI assistance}

During the preparation of this work the authors used ChatGPT in order to improve language and readability. After using this tool, the authors reviewed and edited the content as needed and take full responsibility for the content of the publication.

%**************************************************************************

%\bibliographystyle{apalike}
\bibliography{sim-spin}

@book{grinstein2007implicit,
  title={Implicit large eddy simulation},
  author={Grinstein, Fernando F and Margolin, Len G and Rider, William J},
  volume={10},
  year={2007},
  publisher={Cambridge university press Cambridge}
}

@article{oliveira2025active,
  title={Active Cooling Device: A Flexible, Lab-Scale Experimental Unit to Develop Spatio-Temporal Temperature Control Strategies},
  author={Oliveira Ferreira, Victor and Mainville, Wiebke and Raymond, Vincent and Lamarre, Jean-Michel and Hamel, Antoine and Vaillant, Mikael and Chioua, Moncef and Blais, Bruno},
  journal={Lab-Scale Experimental Unit to Develop Spatio-Temporal Temperature Control Strategies},
  year={2025}
}

@article{bibeau2023artificial,
  title={Artificial neural network to predict the power number of agitated tanks fed by CFD simulations},
  author={Bibeau, Val{\'e}rie and Barbeau, Lucka and Boffito, Daria Camilla and Blais, Bruno},
  journal={The Canadian Journal of Chemical Engineering},
  volume={101},
  number={10},
  pages={5992--6002},
  year={2023},
  publisher={Wiley Online Library}
}

@article{mao2016image,
  title={Image restoration using very deep convolutional encoder-decoder networks with symmetric skip connections},
  author={Mao, Xiaojiao and Shen, Chunhua and Yang, Yu-Bin},
  journal={Advances in neural information processing systems},
  volume={29},
  year={2016}
}

@article{20243817071309 ,
language = {English},
copyright = {Compilation and indexing terms, Copyright 2026 Elsevier Inc.},
copyright = {Compendex},
title = {Enhancing PVT/air system performance with impinging jet and porous media: A computational approach with machine learning predictions},
journal = {Applied Energy},
author = {Farahani, Somayeh Davoodabadi and Zare, Mehdi Khademi and Alizadeh, As'ad},
volume = {377},
year = {2025},
issn = {03062619},
note = {Electrical efficiency;Impinging jet;Layer thickness;Machine-learning;Photovoltaic systems;Porous layers;Porous medium;PV;Solar cell temperatures;Thermal-efficiency;},
URL = {http://dx.doi.org/10.1016/j.apenergy.2024.124509},
}

@article{20252218537271 ,
language = {English},
copyright = {Compilation and indexing terms, Copyright 2026 Elsevier Inc.},
copyright = {Compendex},
title = {Optimizing heat transfer in turbulent impinging jets array: A computational fluid dynamics investigation and machine learning approach for predicting the average Nusselt number},
journal = {Physics of Fluids},
author = {Hnaien, Nidhal and Neffati, Syrine and Abdullah, Nermeen and Hassen, Walid and Aoudia, Mouloud and Kolsi, Lioua},
volume = {37},
number = {5},
year = {2025},
issn = {10706631},
key = {Linear regression},
URL = {http://dx.doi.org/10.1063/5.0275241},
}

@article{20244217219667 ,
language = {English},
copyright = {Compilation and indexing terms, Copyright 2026 Elsevier Inc.},
copyright = {Compendex},
title = {Experimental and machine learning study on the influence of nanoparticle size and pulsating flow on heat transfer performance in nanofluid-jet impingement cooling},
journal = {Applied Thermal Engineering},
author = {Atofarati, Emmanuel O. and Sharifpur, Mohsen and Huan, Zhongjie and Awe, Olushina Olawale and Meyer, Josua P.},
volume = {258},
year = {2025},
issn = {13594311},
key = {Reynolds number},
URL = {http://dx.doi.org/10.1016/j.applthermaleng.2024.124631},
}

@misc{vaillant_2026_18157226,
  author       = {Vaillant, Mikael},
  title        = {Surrogate Model for Heat Transfer Prediction in
                   Impinging Jet Arrays using Dynamic Inlet/Outlet
                   and Flow Rate Control
                  },
  month        = jan,
  year         = 2026,
  publisher    = {Zenodo},
  doi          = {10.5281/zenodo.18157226},
  url          = {https://doi.org/10.5281/zenodo.18157226},
}

@article{tian2020deep,
  title={Deep learning on image denoising: An overview},
  author={Tian, Chunwei and Fei, Lunke and Zheng, Wenxian and Xu, Yong and Zuo, Wangmeng and Lin, Chia-Wen},
  journal={Neural Networks},
  volume={131},
  pages={251--275},
  year={2020},
  publisher={Elsevier}
}

@InProceedings{Arndt_outflow_bc,
author="Arndt, Daniel
and Braack, Malte
and Lube, Gert",
editor="Karas{\"o}zen, B{\"u}lent
and Manguo{\u{g}}lu, Murat
and Tezer-Sezgin, M{\"u}nevver
and G{\"o}ktepe, Serdar
and U{\u{g}}ur, {\"O}m{\"u}r",
title="Finite Elements for the Navier-Stokes Problem with Outflow Condition",
booktitle="Numerical Mathematics and Advanced Applications  ENUMATH 2015",
year="2016",
publisher="Springer International Publishing",
address="Cham",
pages="95--103",
isbn="978-3-319-39929-4"
}

@misc{pytorch,
      title={PyTorch: An Imperative Style, High-Performance Deep Learning Library}, 
      author={Adam Paszke and Sam Gross and Francisco Massa and Adam Lerer and James Bradbury and Gregory Chanan and Trevor Killeen and Zeming Lin and Natalia Gimelshein and Luca Antiga and Alban Desmaison and Andreas Köpf and Edward Yang and Zach DeVito and Martin Raison and Alykhan Tejani and Sasank Chilamkurthy and Benoit Steiner and Lu Fang and Junjie Bai and Soumith Chintala},
      year={2019},
      eprint={1912.01703},
      archivePrefix={arXiv},
      primaryClass={cs.LG},
      url={https://arxiv.org/abs/1912.01703}, 
}

@article{Fu_2022,
title = {Effect of an impinging jet on the flow characteristics and thermal performance of mainstream in battery cooling of hybrid electric vehicles},
journal = {International Journal of Heat and Mass Transfer},
volume = {183},
pages = {122206},
year = {2022},
issn = {0017-9310},
doi = {https://doi.org/10.1016/j.ijheatmasstransfer.2021.122206},
url = {https://www.sciencedirect.com/science/article/pii/S0017931021013053},
author = {Jiahong Fu and Yong Li and Zhen Cao and Bengt Sundén and Junqi Bao and Gongnan Xie}
}

@article{razzaghi_adjoint,
title = {Derivation and application of the adjoint method for estimation of both spatially and temporally varying convective heat transfer coefficient},
journal = {Applied Thermal Engineering},
volume = {154},
pages = {63-75},
year = {2019},
issn = {1359-4311},
doi = {https://doi.org/10.1016/j.applthermaleng.2019.03.068},
url = {https://www.sciencedirect.com/science/article/pii/S1359431119301206},
author = {H. Razzaghi and F. Kowsary and M. Ashjaee}
}

@article{zuckerman2006jet,
  title={Jet impingement heat transfer: physics, correlations, and numerical modeling},
  author={Zuckerman, N and Lior, Noam},
  journal={Advances in heat transfer},
  volume={39},
  pages={565--631},
  year={2006},
  publisher={Elsevier}
}

@misc{Lamarre_2023,
	type = {Patent},
	title = {Multi-input, multi-output manifold for thermocontrolled surfaces},
	url = {https://patents.google.com/patent/US20230182361A1/en?oq=US20230182361A1},
	author = {Lamarre, Jean-Michel and Raymond, Vincent},
	month = jun,
	year = {2023},
}

@inproceedings{weigand2009multiple,
  title={Multiple jet impingement- a review},
  author={Weigand, Bernhard and Spring, Sebastian},
  booktitle={TURBINE-09. Proceedings of international symposium on heat transfer in gas turbine systems},
  year={2009},
  organization={Begel House Inc.}
}

@article{Zhao23022005,
    author = {Wennan Zhao, Kuruchi Kumar and Arun S. Mujumdar},
    title = {Impingement Heat Transfer for a Cluster of Laminar Impinging Jets Issuing from Noncircular Nozzles},
    journal = {Drying Technology},
    volume = {23},
    number = {1-2},
    pages = {105--130},
    year = {2005},
    publisher = {Taylor \& Francis},
    doi = {10.1081/DRT-200047877},
    URL = {https://doi.org/10.1081/DRT-200047877},
    eprint = {https://doi.org/10.1081/DRT-200047877  }
}

@INPROCEEDINGS{deconvolution,
  author={Zeiler, Matthew D. and Krishnan, Dilip and Taylor, Graham W. and Fergus, Rob},
  booktitle={2010 IEEE Computer Society Conference on Computer Vision and Pattern Recognition}, 
  title={Deconvolutional networks}, 
  year={2010},
  volume={},
  number={},
  pages={2528-2535},
  doi={10.1109/CVPR.2010.5539957}}

@incollection{tezduyar1992,
author = {Tezduyar, Tayfun E},
booktitle = {Adv. Appl. Math.},
isbn = {0120020289},
pages = {1--44},
publisher = {Elsevier},
title = {{Stabilized finite element formulations for incompressible flow computations}},
volume = {28},
year = {1991},
url = {https://doi.org/10.1016/S0065-2156(08)70153-4}
}

@misc{PrietoSaavedra2024,
archivePrefix = {SSRN},
author = {Prieto Saavedra, Laura and Munch, Peter and Blais, Bruno},
url = {https://doi.org/10.2139/ssrn.4981567},
title = {{A matrix-free stabilized solver for the incompressible {Navier-Stokes} equations}},
volume = {203},
year = {2024}
}

@article{saavedra2024implicit,
  title={An implicit large-eddy simulation perspective on the flow over periodic hills},
  author={Saavedra, Laura Prieto and Radburn, Catherine E Niamh and Collard-Daigneault, Audrey and Blais, Bruno},
  journal = {Comput. Fluids},
  volume={283},
  pages={106390},
  year={2024},
  publisher={Elsevier}
}

@article{jet_array_correlation,
    author = {Kercher, D. M. and Tabakoff, W.},
    title = {Heat Transfer by a Square Array of Round Air Jets Impinging Perpendicular to a Flat Surface Including the Effect of Spent Air},
    journal = {Journal of Engineering for Power},
    volume = {92},
    number = {1},
    pages = {73-82},
    year = {1970},
    month = {01},
    issn = {0022-0825},
    doi = {10.1115/1.3445306},
    url = {https://doi.org/10.1115/1.3445306},
    eprint = {https://asmedigitalcollection.asme.org/gasturbinespower/article-pdf/92/1/73/5885907/73\_1.pdf},
}

@Article{cross_validation,
AUTHOR = {Allgaier, Johannes and Pryss, Rüdiger},
TITLE = {Cross-Validation Visualized: A Narrative Guide to Advanced Methods},
JOURNAL = {Machine Learning and Knowledge Extraction},
VOLUME = {6},
YEAR = {2024},
NUMBER = {2},
PAGES = {1378--1388},
URL = {https://www.mdpi.com/2504-4990/6/2/65},
ISSN = {2504-4990},
DOI = {10.3390/make6020065}
}

@article{otero2021high,
  title={High-fidelity simulations of multi-jet impingement cooling flows},
  author={Otero-P{\'e}rez, J Javier and Sandberg, Richard D and Mizukami, Satoshi and Tanimoto, Koichi},
  journal={Journal of Turbomachinery},
  volume={143},
  number={8},
  pages={081011},
  year={2021},
  publisher={American Society of Mechanical Engineers}
}

@article{uddin2024heat,
  title={Heat transfer by jet impingement: A review of heat transfer correlations and high-fidelity simulations},
  author={Uddin, Naseem and Kee, Priscilla Tang Wei and Weigand, Bernhard},
  journal={Applied Thermal Engineering},
  pages={124258},
  year={2024},
  publisher={Elsevier}
}

@article{barbosa2023convection,
  title={Convection from multiple air jet impingement-A review},
  author={Barbosa, Fl{\'a}via V and Teixeira, Senhorinha FCF and Teixeira, Jos{\'e} CF},
  journal={Applied Thermal Engineering},
  volume={218},
  pages={119307},
  year={2023},
  publisher={Elsevier}
}

@article{fujimori2024dns,
  title={DNS analysis of the effect of control parameters on the heat transfer performance of intermittently controlled impinging jets},
  author={Fujimori, Koki and Tsujimoto, Koichi and Ando, Toshitake and Takahashi, Mamoru},
  journal={International Journal of Heat and Fluid Flow},
  volume={106},
  pages={109301},
  year={2024},
  publisher={Elsevier}
}

@article{hopmann2020development,
  title={Development of a novel control strategy for a highly segmented injection mold tempering for inline part warpage control},
  author={Hopmann, Christian and Kahve, Cemi and Schmitz, Mauritius},
  journal={Polymer Engineering \& Science},
  volume={60},
  number={10},
  pages={2428--2438},
  year={2020},
  publisher={Wiley Online Library}
}

@article{chiriac2002numerical,
  title={A numerical study of the unsteady flow and heat transfer in a transitional confined slot jet impinging on an isothermal surface},
  author={Chiriac, Victor A and Ortega, Alfonso},
  journal={International Journal of Heat and Mass Transfer},
  volume={45},
  number={6},
  pages={1237--1248},
  year={2002},
  publisher={Elsevier}
}

@article{plant2023review,
  title={A review of jet impingement cooling},
  author={Plant, Robert D and Friedman, Jacob and Saghir, M Ziad},
  journal={International Journal of Thermofluids},
  volume={17},
  pages={100312},
  year={2023},
  publisher={Elsevier}
}

@article{salavatidezfouli2023deep,
  title={Deep reinforcement learning for the heat transfer control of pulsating impinging jets},
  author={Salavatidezfouli, Sajad and Stabile, Giovanni and Rozza, Gianluigi},
  journal={arXiv preprint arXiv:2309.13955},
  year={2023}
}

@article{salavatidezfouli2024predictive,
  title={A Predictive Surrogate Model for Heat Transfer of an Impinging Jet on a Concave Surface},
  author={Salavatidezfouli, Sajad and Rakhsha, Saeid and Sheidani, Armin and Stabile, Giovanni and Rozza, Gianluigi},
  journal={arXiv preprint arXiv:2402.10641},
  year={2024}
}

@article{singh2021numerical,
  title={Numerical simulations and optimization of impinging jet configuration},
  author={Singh, Alankrita and Chakravarthy, Balaji and Prasad, BVSSS},
  journal={International Journal of Numerical Methods for Heat \& Fluid Flow},
  volume={31},
  number={1},
  pages={1--25},
  year={2021},
  publisher={Emerald Publishing Limited}
}

@article{fawaz2024artificial,
  title={Artificial neural networks application on average and stagnation Nusselt number prediction for impingement cooling of flat plate with helically coiled air jet},
  author={Fawaz, HE and Osama, Mostafa M and Maghrabie, Hussein M},
  journal={Journal of Thermal Science and Engineering Applications},
  volume={16},
  number={2},
  pages={021012},
  year={2024},
  publisher={American Society of Mechanical Engineers}
}

@article{lethe1.0,
  title={Lethe 1.0: An open-source parallel high-order computational fluid dynamics software framework for single and multiphase flows},
  author={Alphonius, Amishga and Barbeau, Lucka and Blais, Bruno and Gaboriault, Olivier and Gu{\'e}vremont, Olivier and Lamouche, Justin and Laurentin, Pierre and Marquis, Oreste and Munch, Peter and Ferreira, Victor Oliveira and others},
  journal={Computer Physics Communications},
  pages={109880},
  year={2025},
  publisher={Elsevier}
}

@article{afzal2017effects,
  title={Effects of Latin hypercube sampling on surrogate modeling and optimization},
  author={Afzal, Arshad and Kim, Kwang-Yong and Seo, Jae-won},
  journal={International Journal of Fluid Machinery and Systems},
  volume={10},
  number={3},
  pages={240--253},
  year={2017},
  publisher={Turbomachinery Society of Japan, Korean Society for Fluid Machinery, Chinese~…}
}

@incollection{martin1977heat,
  title={Heat and mass transfer between impinging gas jets and solid surfaces},
  author={Martin, Holger},
  booktitle={Advances in heat transfer},
  volume={13},
  pages={1--60},
  year={1977},
  publisher={Elsevier}
}

@article{jambunathan1992review,
  title={A review of heat transfer data for single circular jet impingement},
  author={Jambunathan, K and Lai, Eliza and Moss, MAm and Button, BL},
  journal={International journal of heat and fluid flow},
  volume={13},
  number={2},
  pages={106--115},
  year={1992},
  publisher={Elsevier}
}

@article{katti2008experimental,
  title={Experimental study and theoretical analysis of local heat transfer distribution between smooth flat surface and impinging air jet from a circular straight pipe nozzle},
  author={Katti, Vadiraj and Prabhu, SV},
  journal={International Journal of Heat and Mass Transfer},
  volume={51},
  number={17-18},
  pages={4480--4495},
  year={2008},
  publisher={Elsevier}
}

\pagebreak

\appendix

\section{CNN Vs. ANN} \label{sup:cnn_vs_ann}

\begin{figure}[ht!]
    \centering
    \includegraphics[width=0.8\linewidth]{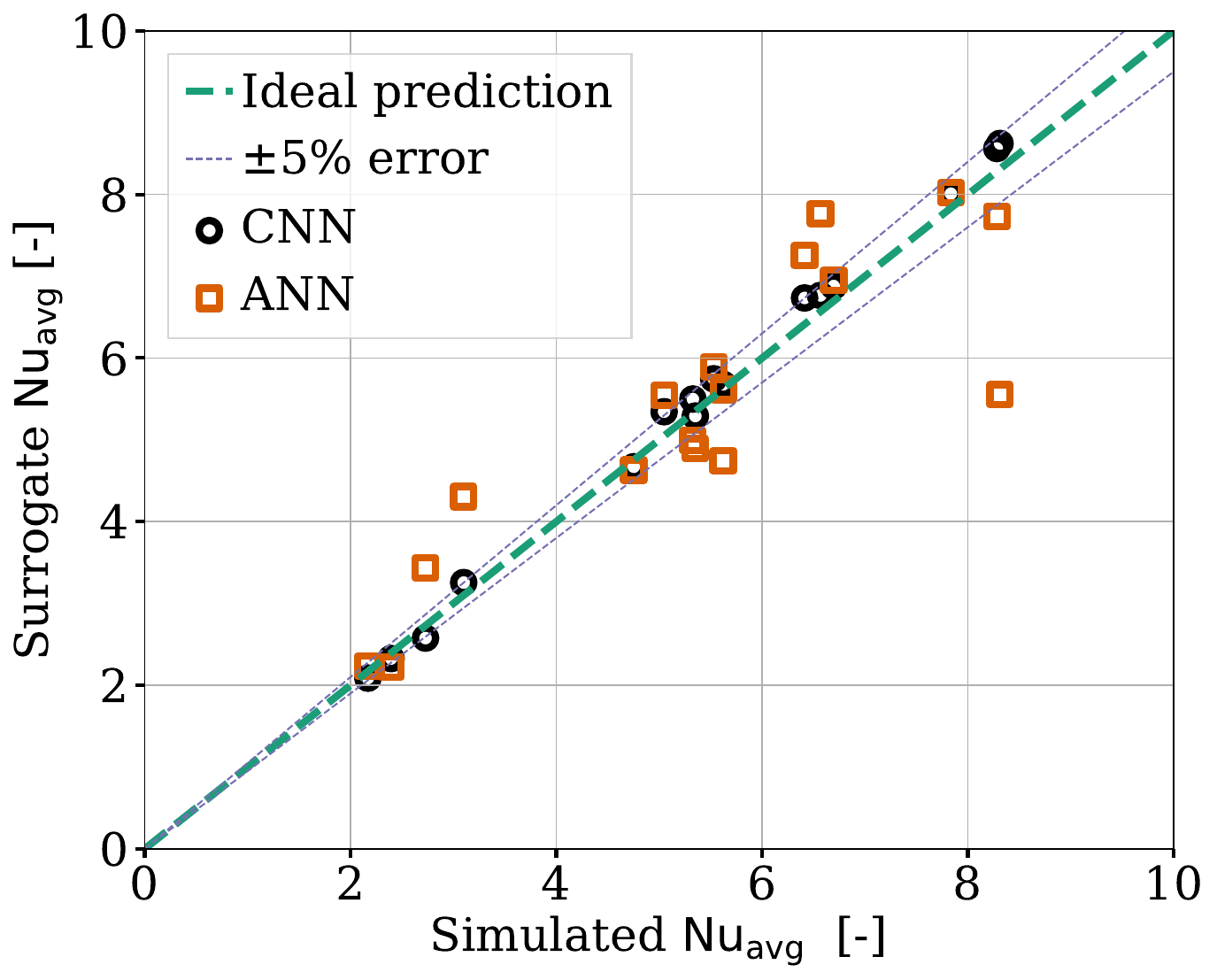}
    \caption{Comparison between a surrogate trained for the 5 by 1 geometry using a CNN-based architecture and using an ANN-based architecture.}
    \label{fig:cnn_vs_ann}
\end{figure}

\pagebreak

\end{document}